\documentclass[%
 aip,
rsi,%
 amsmath,amssymb,
 reprint,%
 longbibliography,
]{revtex4-1}

\usepackage{graphicx}% Include figure files
\usepackage{dcolumn}% Align table columns on decimal point
\usepackage{bm}% bold math
\usepackage{hyperref} % for citing url within the \web custom command, see below
\hypersetup{   colorlinks,   menucolor=blue,   linkcolor=blue,   citecolor=blue,   urlcolor=blue}
\newcommand\web[1]{\href{#1}{website}}

\newcommand{\keff}{k_\mathrm{eff}}
\newcommand{\veckeff}{\vec{k}_\mathrm{eff}}

\newcommand\stxt[1]{_{\text{#1}}} % subscript
\newcommand{\radps}{rad.s$^{-1}$}

\begin{document}

\preprint{AIP/123-QED}

\title[High-accuracy inertial measurements with cold-atom sensors]{High-accuracy inertial measurements with cold-atom sensors}% Force line breaks with \\
%\thanks{Footnote to title of article.}

\author{Remi Geiger}
\email{remi.geiger@obspm.fr}
\author{Arnaud Landragin}
\email{arnaud.landragin@obspm.fr}
\author{S\'ebastien Merlet}
\email{sebastien.merlet@obspm.fr}
\author{Franck Pereira Dos Santos}
\email{franck.pereira@obspm.fr}
\affiliation{ 
LNE-SYRTE, Observatoire de Paris - Universit\'e PSL, CNRS, Sorbonne Universit\'e, 61, avenue de l'Observatoire, 75014 Paris, France
%\\This line break forced with \textbackslash\textbackslash
}%

\date{\today}% It is always \today, today,

\begin{abstract}
The research on cold-atom interferometers gathers a large community of about 50 groups worldwide both in the academic and now in the industrial sectors. The interest in this sub-field of quantum sensing and metrology lies in the large panel of possible applications of cold-atom sensors for measuring inertial and gravitational signals with a high level of stability and accuracy. This review presents the evolution of the field over the last 30 years and focuses on the acceleration of the research effort in the last 10 years.  The article describes the physics principle of cold-atom gravito-inertial sensors as well as the main parts of hardware and the expertise required when starting the design of such sensors. It then reviews the progress in the development of instruments measuring gravitational and inertial signals, with a highlight on the limitations to the performances of the sensors, on their applications, and on the latest directions of research.

%Valid PACS numbers may be entered using the \verb+\pacs{#1}+ command.
\end{abstract}

\keywords{Atom interferometry, cold atoms, inertial sensors, quantum metrology and sensing.}%Use showkeys class option if keyword
\maketitle

%\begin{quotation}
%The ``lead paragraph'' is encapsulated with the \LaTeX\ 
%\verb+quotation+ environment and is formatted as a single paragraph before the first section heading. (The \verb+quotation+ environment reverts to its usual meaning after the first sectioning command.)Note that numbered references are allowed in the lead paragraph.
% The lead paragraph will only be found in an article being prepared for the journal \textit{Chaos}.
%\end{quotation}

\section{\label{sec:intro} Introduction} 
%\textcolor{red}{\textit{Goal of this review: describe to a non specialist the field of cold-atom inertial sensors (CAIS). Focus on studies that perform measurements of inertial quantities: gravimeters, gradiometers, accelerometers, gyroscopes. Goals: understand principle, main sensitivity and accuracy limits that drive the design, understand the hardware, have an overview of the groups working in the field and of the most important results, identify latest developments and applications.\\ Say that we don't present tests of fundamental physics, neither trapped atom interferometers and chip-based technologies (just mention they exist). Give references to other reviews in that context. Book by Berman on atom interefrometry \cite{berman1997atom}. Book of 2014 with proceedings of Enrico Fermi school \cite{EnricoFermiBook2014}. List of other reviews on the topic, briefly say what they explain and what is missing to motivate the present review. Review "Search for new physics with atoms and molecules" \cite{SafronovaRMP2018} by Safronova et al. Barrett et al \cite{Barrett2016review}; Robins et al \cite{Robins2013} (application of atom lasers for precision inertial measurements); Cronin et al \cite{Cronin2009} (interferometry with atoms and molecules).   Review on taking atom interferometers out of the field by Bongs et al\cite{Bongs2019.}}}

Interferometry with matter waves nearly dates back to the first ages of quantum mechanics as the concept of matter waves played a key role in the development of the quantum theory, following the theoretical work of de Broglie in 1924 and the ensuing experiments of Davisson, Germer and Thomson with electron beams. Since then, performing interference experiments with various types of matter-waves has driven the efforts of  several communities working with electrons, neutrons, atoms, molecules, or anti-matter.
The field of atom interferometry has developped rapidly with the advancement of atomic physics, which offers a high level of control and reliability to the experimental physicist. This degree of control has become even more impressive since the advent of laser cooling techniques in the 1980s, which enhance the wave nature of atoms by increasing their coherence length. 

Since  pioneering experiments in 1991 \cite{Kasevich1991,Riehle1991,Keith1991,carnal_youngs_1991,robert_atomic_1991}, the field of atom interferometry has constantly grown, with an acceleration in the last 10 years.
Cold-atom inertial sensors based on light-pulse atom interferometry have reached sensitivity and accuracy levels competing with or beating inertial sensors based on different technologies. 
Such sensors cover various applications ranging from geophysics and inertial sensing to metrology and tests of fundamental physics. Addressing these applications requires to constantly push further the performances of quantum sensors.

As of 2020, about 50 research groups worldwide are actively developing atom interferometers for different applications, and investigating techniques to improve the performances of cold-atom inertial sensors. 
Currently, the  research focuses on three mains aspects:
\begin{enumerate}
\item pushing the performances of current  sensors;
\item identifying new sensor architectures or  generic techniques  that can bring performance improvement or simplified architectures;
\item using atom interferometers for various fundamental and/or field applications.
\end{enumerate}

Improving the performances of atom inertial sensors covers different aspects: their sensitivity, but also their stability, accuracy, dynamic range, compactness, transportability, ease-of-use and cost.  While the first 20 years of research were essentially focused on sensitivity  improvements and tests of fundamental physics in laboratory environments, more  projects have started to address field applications. In particular, this is the case for inertial guidance, which requires at the same time high levels of stability, wide dynamic ranges and high sampling frequencies, compactness and robustness. For this field of application though, cold-atom sensors are not yet mature enough to compete with other technologies in all these aspects (e.g. ring laser gyroscopes for navigation).  In that sense, the course for greater performance is, for example, at the core of the \textit{Quantum Sensors and Metrology} pillar of the several quantum technology programs over the world.

Several reviews of the field have been published in the last 10 years: the review Ref.~\cite{Cronin2009} in 2009 presents the whole field of matter-wave interferometry and detailed some of the  cold-atom  inertial sensor developments; more recently, Ref.~\cite{Barrett2014} from 2014 presents the advancements related to atomic gyroscopes, and Ref.~\cite{Barrett2016review} from 2016 presents the principle of inertial quantum sensors using light and matter and shows some examples; a perspective (Ref.~\cite{Bongs2019}) published in 2019 presents the challenges required to bring atom interferometers out of the laboratory; the review in Ref.~\cite{SafronovaRMP2018} from 2018  presents in details the application of cold-atom sensors to tests of fundamental physics and search for new physics. At a more specialized level, some review articles address specific problems linked to cold-atom sensors (e.g. the prospect of using atom-lasers as a source for atom interferometers\cite{Robins2013}), or  specific applications (e.g. gravitational wave detection by atom interferometry \cite{Geiger2016,Loriani2019}).
 Two books, published in 1997 (Ref.~\cite{berman1997atom}) and  2014 (Ref.~\cite{Tino2014}) gathering specialized contributions from experts in the field allow to catch in more details the various techniques and applications. To avoid overlap with these contributions and address a general  audience, we focus here on cold-atom sensors aiming at measuring inertial signals, with the aim to present an exhaustive and up-to-date view of the field, including both physical and system engineering aspects.
 Trapped atom interferometers, which represent an interesting perspective for both fundamental studies and miniaturized sensors (but yet not competitive in terms of sensitivity and accuracy) are also not described in details here (see for instance the recent review of Ref.~\cite{GarridoAlzar2019}).

This article is organized as follows. 
In section \ref{sec:principle}, we present the principles of light-pulse atom interferometry which are generic to the different sensor architectures described in this review. We explain the main limitations to the sensitivity which drive the design of instruments. 
Section \ref{sec:hardwave} presents the most important elements of hardware common to any cold-atom inertial sensor. 
Section \ref{sec:gravity_sensors} shows the developments of cold-atom gravimetry and its applications. 
Section \ref{sec:inertial_sensors} focuses on the research on accelerometers and gyroscopes as the potential building blocks of future inertial navigation systems. 
Section \ref{sec:other_measurements} briefly presents for completeness an overview of other inertial measurements performed with cold-atom sensors, such as measurement of recoil velocities, prospects for gravitational wave detection or tests of the weak equivalence principle.
Section \ref{sec:new_techniques} describes the latest atomic physics techniques under study in academic laboratories to improve the performances of cold-atom inertial sensors.
After the conclusion of the review, a list of summary points highlights the most important ideas of the article and  appendix \ref{sec:list_groups} presents an up-to-date list of the  different research groups actively working in the field of cold-atom inertial sensors.

\section{\label{sec:principle} Principle}
%\textcolor{red}{\textit{Explain the basic principle of light-pulse interferometry with cold atoms. Focus on Raman transitions but mention that Bragg exist. Explain the atom optics from the light pulses, explain where the phase comes from. Then explain sensitivity limit and why cold atoms.}}

In this section, we explain the basic principle of light pulse interferometry with cold atoms, from the description of the light pulse beamsplitters to the creation of an atom interferometer. 

We consider here the case of beamsplitters based on two photon transitions, such as based on stimulated Raman transitions or Bragg diffraction ; the later being a degenerated case of the former. It represents the vast majority of atom interferometers, because it allows at the same time for very high sensitivities and accuracies, for a good compromise in terms of simplicity. Indeed, the use of optical transitions allows for both a large velocity transfer to the atom (of the order of the cm.s$^{-1}$) needed for the sensitivity, and a very good control of the diffraction process, required for the accuracy. Furthermore, the use of two-photon transitions releases the constraint on the control of the optical phase of the lasers used in the beamsplitters, as only the phase difference between the two laser beams need to be controlled at first sight.

\begin{figure*}[t]
	\centering
	\includegraphics[width=\linewidth]{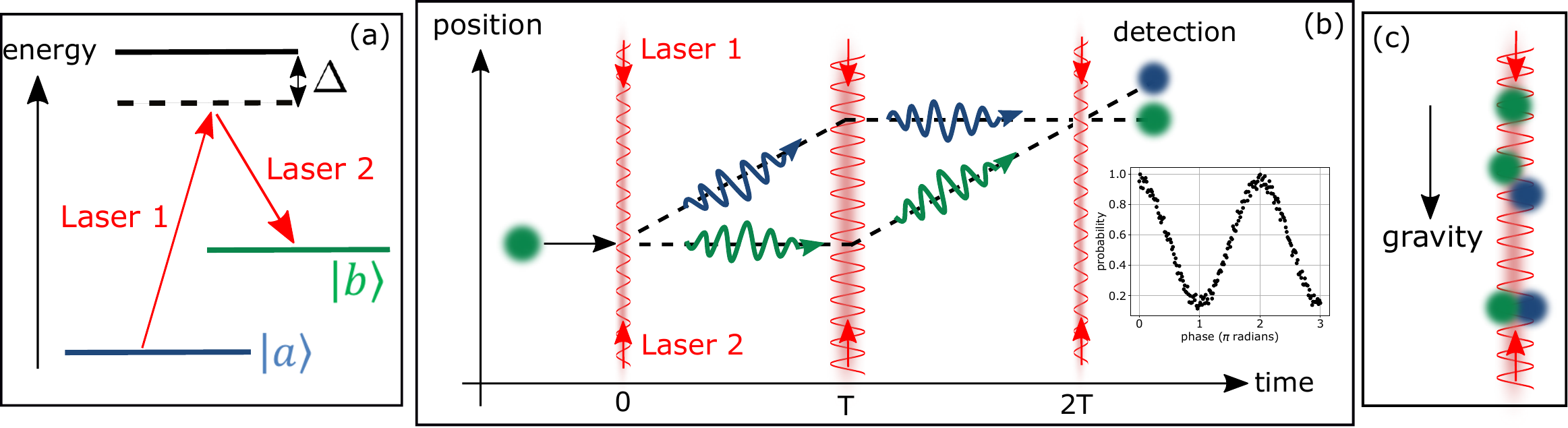}
	\caption{\textbf{Principle of a light-pulse cold-atom inertial sensor}. (a) Three-level atom coupled to two counter-propagating laser beams. The atom is subject to a stimulated two-photon process by absorption of a photon from laser 1 and stimulated emission of a photon in the mode of laser 2. This level diagram is typical of alkali atoms with two hyperfine ground states and an excited state manifold from which the two lasers are detuned in frequency by $\Delta$.  
	The transition between internal states is accompanied by a change of momentum given by $\hbar (\vec{k}_1-\vec{k}_2)\equiv \hbar \veckeff$. (b) A sequence of three light pulses allows to split, deflect and recombine the atomic waves to form an atomic Mach-Zehnder interferometer. The detection of the atom state at the output yields the atomic interference which is modulated by the difference of phase along the two arms.  (c) Example of arrangement of the laser beams in the vertical direction in which atoms are free falling. }
	\label{fig:principle}
\end{figure*}

In the case of Raman transitions, the atoms interact with two counter-propagating lasers, of angular frequencies $\omega_1$ and $\omega_2$ and wavevectors $k_1$ and $k_2$. These two light fields are detuned from a strong electronic transition (typically a few hundreds of MHz to a few GHz away from the D2 line of alkali atoms) but their frequency difference matches the energy difference between two fundamental atomic states $|a\rangle$ and $|b\rangle$, which are then coupled by the light fields in a so-called lambda scheme (Fig.\ref{fig:principle}.a). Atoms initially in the state $|a\rangle$ will absorb a photon in the laser 1 and de-excite by stimulated emission of a photon in the laser 2, ending up in the state $|b\rangle$. Conservation of momentum implies that the two coupled states differ in momentum by the momentum transfer $\hbar(\vec{k_1}-\vec{k_2})=\hbar \vec{k}_\text{eff}$. The process being coherent, the system undergoes Rabi oscillations, such that by adjusting the duration and Rabi frequency of the laser pulses, one can prepare the state of an atom in a superposition of the two couple states with controlled weights. In particular, a so-called $\pi/2$ pulse acts as a matter wave beamsplitter, placing an atom in a balanced 50/50 superposition of the two couple states. A twice longer pulse is a $\pi$ pulse, which swaps the two states, acting as a mirror for the matter wave.   

In the most simple case, a sequence of three $\pi/2-\pi-\pi/2$ pulses (of duration $\tau-2\tau-\tau$ for a constant Rabi frequency), separated by free evolutions times $T-T$ then realizes the atom interferometer displayed in figure~\ref{fig:principle}.b. There, the three pulses act as beamsplitters and mirrors, separating, redirecting and recombining the two partial wavepackets. This interferometer geometry is most often referred to as a Mach Zehnder interferometer due to its analogy of the latter optical interferometer. 

The populations in the two output ports of the interferometer are measured using a state selective fluorescence detection~\cite{Borde1989}. One finally derives out of these two populations ($N1, N2$) the transition probability $P= N1/(N1+N2)$. As in any other two-wave interferometer, it is given by $P=P_0+C/2\times \text{cos}(\Delta\Phi)$, where $C$ is the interferometer contrast and $\Delta\Phi$ the interferometer phase. This phase is the difference between the phases accumulated by the atomic wavepackets along the two arms of the interferometer. 

At the laser pulses, the phase difference between the counter-propagating lasers $\phi$ gets imprinted onto the atomic wavefunctions, so that in the end, the interferometer phase shift is given by a linear combination of the lasers phase difference $\phi$ at the three pulses\cite{Storey1994,Borde2004}: 
\begin{equation}
\Delta\Phi=\phi_1-2\phi_2+\phi_3.
\end{equation}
For free falling atoms, this leads to 
\begin{equation}
\Delta\Phi=-\veckeff\cdot \vec{a} T^2 + \veckeff\cdot(2\vec{\Omega}\times \vec{v}) T^2,
\label{eq:phase_shift_accel_rot}
\end{equation}
where $\vec{a}$ and $\vec{\Omega}$ are respectively the acceleration and the rotation rate of the experiment with respect to a reference frame defined by the purely inertial motion of the atoms. This dependence to inertial forces allows one to actually realize sensitive and absolute atom interferometry based inertial sensors: accelerometers and gyroscopes. 

The intrinsic sensitivity of these sensors is limited by the noise on the measurement of the transition probability, and ultimately by the so-called quantum projection noise resulting from the projective measurements of the populations in the two ports of the interferometer \cite{ itano_quantum_1993}.

\section{\label{sec:hardwave} System engineering}
\begin{figure}[t]
	\centering
	\includegraphics[width=\linewidth]{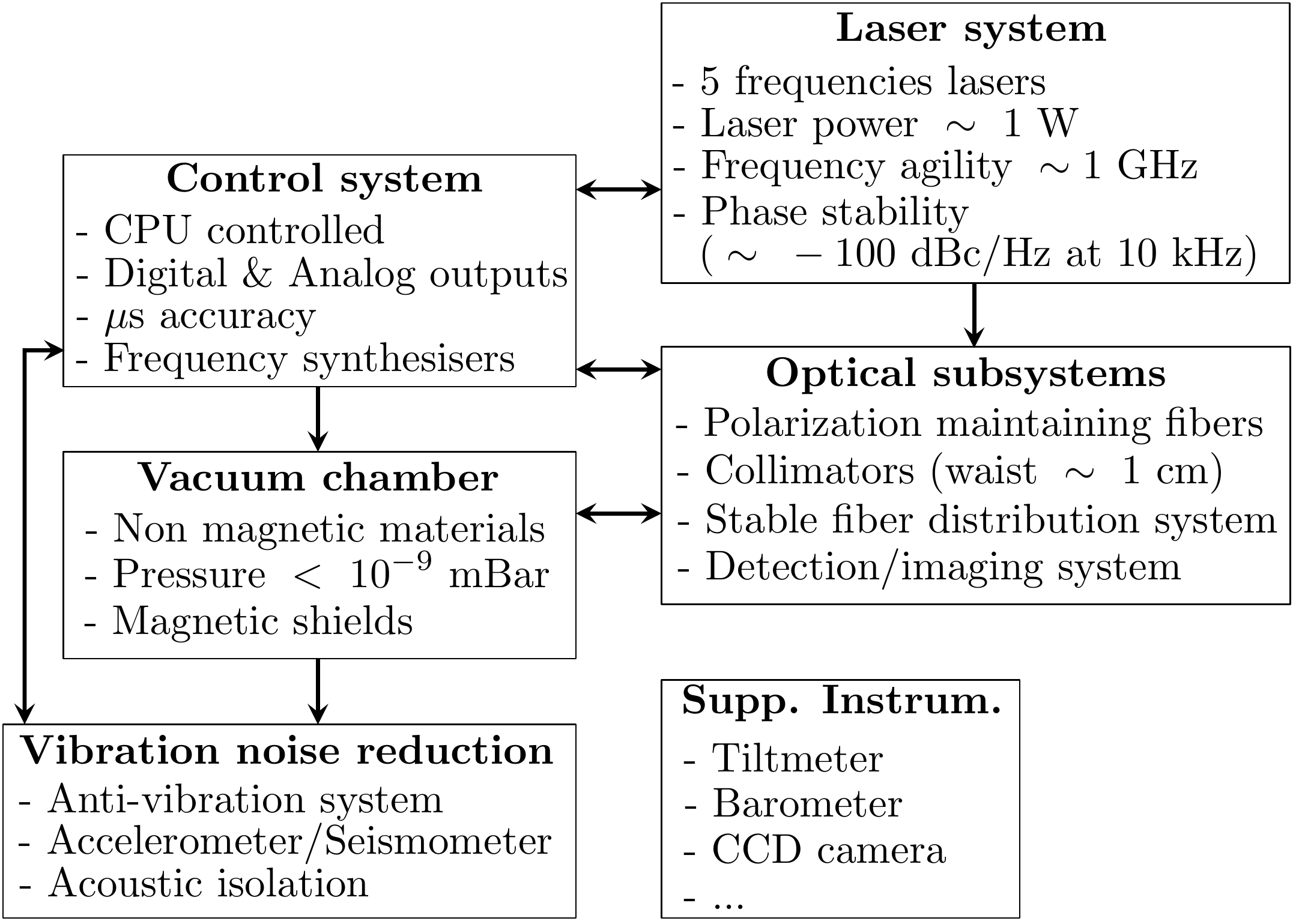}
	\caption{\textbf{Overview of the systems required in a cold-atom inertial sensor.}} 
	\label{fig:system_engineering}
\end{figure}

The hardware common to all cold-atom inertial sensors consists of a vacuum chamber where the atoms are interrogated, a laser system required for cooling, manipulating and detecting them, an automatized control system to operate and interface the instruments, and some auxiliary instrumentation to stabilize the experiment. A general view of the different sub-systems is presented in figure~\ref{fig:system_engineering}. 

%\begin{figure*}[t]
%	\centering
%	\includegraphics[width=\linewidth]{system_engineering.png}
%	\caption{\textbf{Overview of the systems required in a cold-atom inertial sensor.}}
%	\label{fig:system_engineering}
%\end{figure*}

\subsection{Vacuum system and cold-atom source}

%\textcolor{red}{explain the source (dispenser or cell, 2D MOT sometimes, level of vacuum required, non magnetic materials, optical windows required. Pumps.}

At first, a sample of cold atoms is prepared in an Ultra-High Vacuum (UHV) chamber surrounded by magnetic shields, using standard laser cooling methods. The level of vacuum in the chamber must be below $10^{-9}$~hPa in order to be non limiting for the coherence of the system with atoms evolving freely during hundreds of ms. Such UHV level is reached with combinations of pumping technologies such as turbomolecular pumps during backing of the chamber, and getter and ion pumps after baking. Moreover, the vacuum chamber shall be made of non-magnetic materials (e.g. Titanium, aluminium, glass ...) in order to limit the magnetic field gradients which are a source of stray forces owing to the second order Zeeman effect (for atom interferometers operating on transitions insensitive to the first order Zeeman effect). When metallic, the chambers are machined to accommodate typically a dozen of optical windows. They are interfaced with coils that generate magnetic fields: a magnetic gradient for the MOT phase and an homogeneous bias field for the interferometer itself. Most often alkali atoms are used, and a preferred choice is $^{87}$Rb (interrogation wavelength on the $D_2$ line $\lambda=780$~nm (Ref.\cite{Steck})). Loading of a 3 dimensional Magneto Optical Trap directly from a background vapor \cite{ monroe_very_1990} or the intense flux of a 2D MOT\cite{ dieckmann_two-dimensional_1998} allows for gathering of order of $10^8$ atoms in 100~ms. Deep molasses cooling allows reaching temperature close to the recoil limit of the order of 2~$\mu$K. Atoms are then launched upwards in a fountain geometry \cite{ kasevich_rf_1989}, or simply released in free fall from the molasses \cite{LeGouet2008}. A sequence of microwave, pusher, and eventually Raman, pulses is then used to prepare the atoms in a pure Zeeman insensitive $m_F=0$ state, eventually with a narrower velocity distribution. This preparation phase reduces the sensitivity of the source to stray magnetic field fluctuations and increases the contrast of the interferometer, which is in general limited by the finite velocity spread of the source.

\subsection{Laser system}
%\textcolor{red}{Show an example of laser system to illustrate the various outputs required and give some references on the different technologies (telecom doubled versus semiconductor).}
Since cooling the atoms and manipulating their quantum state is performed with lasers, the optical system represents a key subsystem of a cold-atom inertial sensor. The choice of laser technology is intimately linked with the nature of the atom used. Since alkali atoms are widely used, in particular Rubidium and Cesium interrogated on their $D_2$ lines (respectively at 780~nm and 852~nm), semiconductor diode laser technology has been historically vastly deployed \cite{baillard2006}. But telecom based laser sources have also attracted a lot of attention owing to the presence of qualified components (e.g. for field or space applications). This technology leads to commercially available laser systems for cold-atom inertial sensor experiments. To get enough optical power, of order of hundreds of mW, fiber or semiconductor amplifiers are used. All lasers need to be precisely tuned onto specific frequencies. This is realized using a number of frequency locking techniques: saturation spectroscopy in vapour cells, offset locks based on beatnote and acousto-optic modulation. In addition, Raman lasers need to be phase locked together.

\begin{figure}[t]
	\centering
	\includegraphics[width=\linewidth]{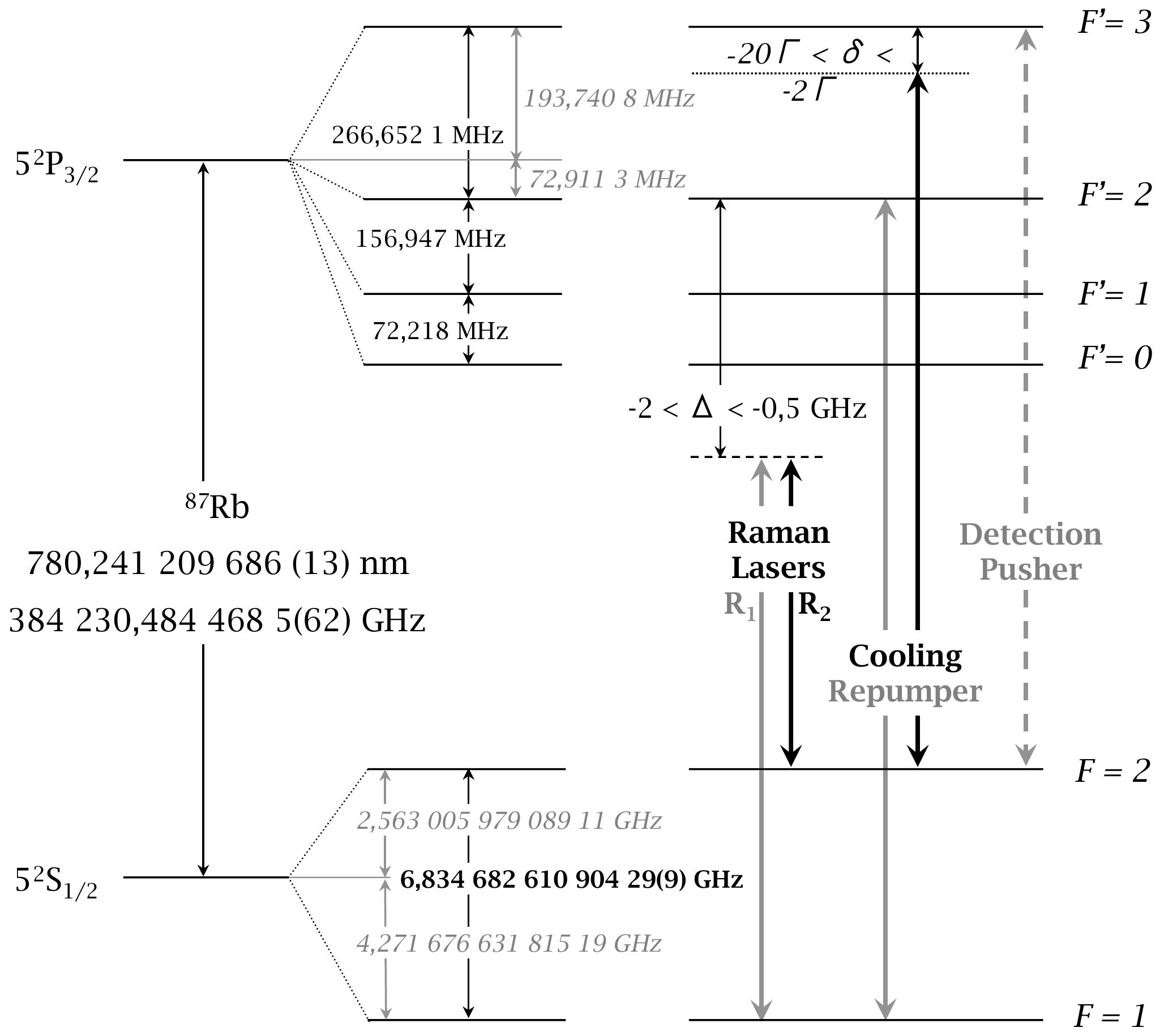}
	\caption{\textbf{Level diagram of the $^{87}$Rb $D_2$ line}. (from Ref.\cite{Steck}) The five laser frequencies required for cooling, detecting and manipulating the atoms with a stimulated two-photon Raman process are represented. 
  }
	\label{fig:niveaux}
\end{figure}

\begin{figure}[t]
	\centering
	\includegraphics[width=\linewidth]{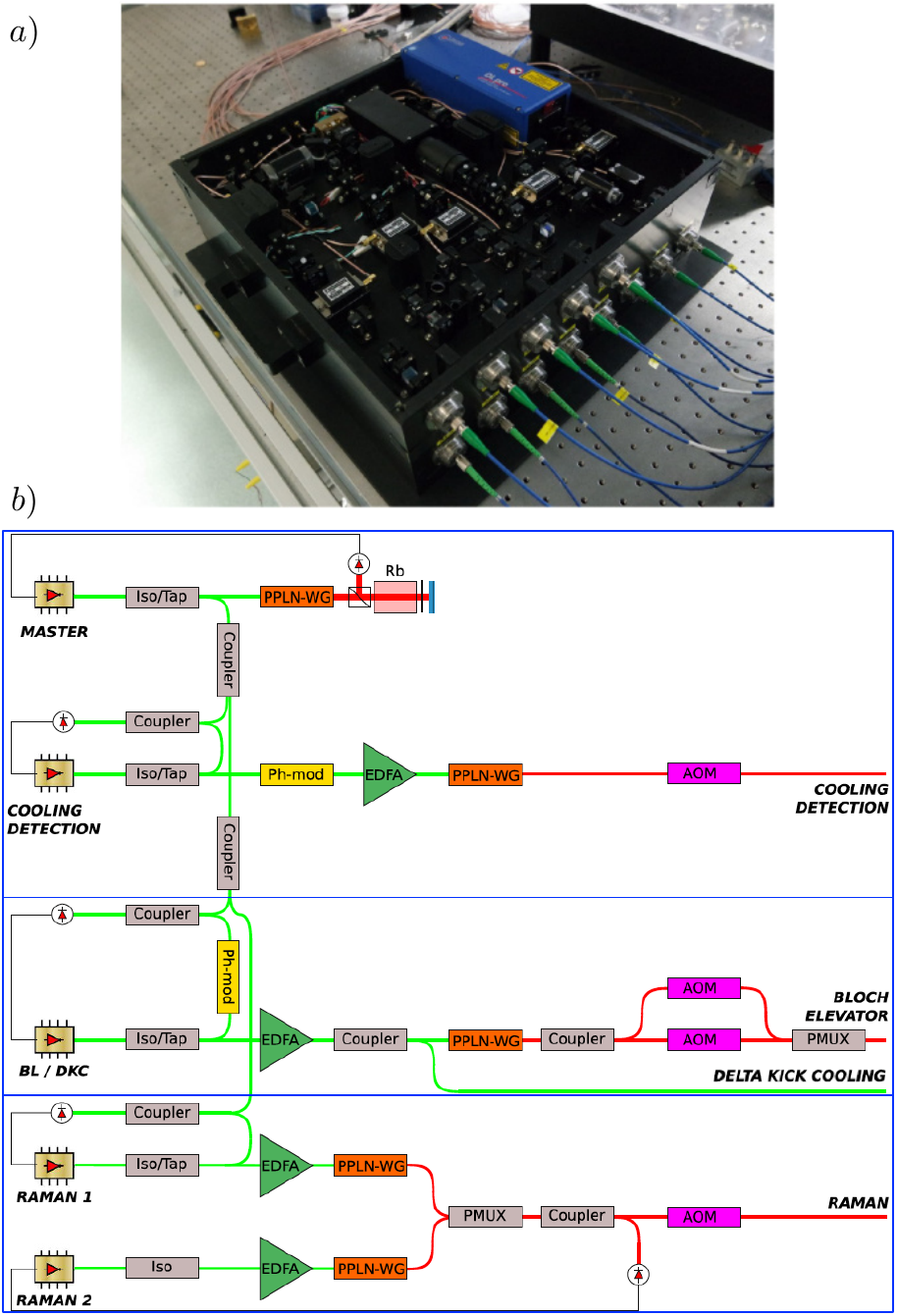}
\caption{\textbf{Examples of laser system.} a) Photography of a free space compact laser system. (Adapted with permission from X. Zhang \textit{et al.} "Compact portable laser system for mobile cold atom gravimeters", Appl. Opt. \textbf{57}, 6545-6551 (2018) \copyright The Optical Society (Ref.~\cite{Zhang2018})) . b) Example of optical architecture for an industrial telecom-doubled based system; Iso/Tap: optical isolator with tap coupler, PPLN-WG: waveguide PPLN crystal, Rb: Rubidium cell, Ph-mod: phase modulator, EDFA: Erbium-Doped Fiber Amplifier, AOM: Acousto-Optic Modulator, PMUX: polarization multiplexer. The upper part shows the master laser, followed by the cooling/detection laser. The bottom part shows the Raman lasers. The middle panel presents an optional system which allows to realize a Bloch elevator to launch the atoms and a delta kick collimator.   (Adapted by permission from Springer Nature Customer Service Centre GmbH : Springer Nature, Eur. Phys. J. D, "A prototype industrial laser system for cold atom inertial sensing in space", R. Caldani \textit{et al.} \copyright (2020), (Ref.~\cite{Caldani2019})).}
	\label{fig:laser_system}
\end{figure}

The five optical frequencies needed to perform a cold-atom inertial sensor are represented in figure~\ref{fig:niveaux}, for $^{87}$Rb interferometers based on Raman transitions. A variety of different laser systems have been developed and published, with different number of lasers, ranging from five to only one, with designs constrained by the size, the final application, the measurement environment conditions and the evolution of technologies~\cite{cheinet_compact_2006, Muller2006, stern_light-pulse_2009, Carraz2009, schmidt_portable_2011, Hu2013, merlet2014, Leveque2014, Leveque2015, schkolnik_compact_2016, Wu2017, Zhang2018, zi2019, Caldani2019}. Figure~\ref{fig:laser_system} displays a compact free space optical system and a complete architecture of a fibered optical bench which reached a Technology Readiness Level (TRL) of 4.

\subsection{Optical subsystems}

Optical collimators are needed to shape and route light beams from the laser system to the vacuum chamber. When only one collimator was used in Ref.~\cite{bodart_cold_2010} to realize all the functions of a gravimeter (trapping, interferometer and detection), most of the sensors actually use a ten of collimators. When driving Raman transitions, the phase difference between the Raman lasers needs to be stable and well controlled, not only in time (which requires the two beams to be phase-locked), but also in space (same wavefront for both counterpropagating laser beams). It can be realized by shaping both beams together using a single common collimator and retroreflecting them on a common mirror). Special care has thus to be paid in the realization of this Raman collimator: most often fibered for better stability in direction and shape, with a large beam waist for flat wavefronts (see part.\ref{sectionAccuracy}), with a well defined polarisation to maximize the coupling to the atoms and an intensity profile as flat as possible in order to keep the coupling homogeneous all along the atom trajectory. The retroreflecting system, usually composed of a quarter-wave plate and a mirror, constitutes a key subsystem in order to achieve high accuracy, as it defines the lasers equiphase, which constitute the reference for the acceleration measurement. A very high level of optical quality (in particular planeity) is thus required for this system (see part.\ref{sectionAccuracy}) in order to reduce the variation of phase difference between the two counterpropagating beams sampled at the three pulses. To keep any eventual residual bias due to such wavefront distortions stable, one needs stable atom trajectories. These are determined by the cloud temperature, its initial mean velocity and position in the beams, which can be stabilized by locking the optical powers of the molasses beams\cite{Farah2014PRA}. Finally, the detection system which collects the fluorescence signals from the clouds needs to be homogeneous and symmetric in order to avoid asymmetry effects\cite{Farah2014PRA, gillot_limits_2016}.

\subsection{Control system}
The automatized control of the apparatus, in particular of the laser system, shares the same constraints as typical atomic physics experiments: it requires tens of analog and digital outputs with temporal resolution below the $\mu$s, agile frequency synthesizers (RF and microwaves) and analog-to-digital converters. An important literature is available on the topic which is a key concern in experiments due to the rapid evolution of hardware control standards as well as operating system versions. We therefore refer the reader to the references in the most recent publications or online materials on this subject, e.g. Refs. ~\cite{Malek2019,Bertoldi2020HAL,ARTIQwebsite}.

\subsection{Vibration noise reduction}\label{sec:noise}
When increasing the interrogation time $T$, phase noise induced by parasitic vibrations (i.e. at frequencies higher that the sampling frequency) rapidly constitutes the dominant limit to the sensitivity. Already when $2T$ gets larger than a few ms, ground vibration noise will typically limit the sensitivity, to a level of $\sim 10^{-5}~$m.s$^{-2}$.$\tau^{-1/2}$, so that the sensors need to be isolated and/or the vibrations measured to reject them. Different methods have been developed in order to reduce this noise source, which have to be adapted to the environmental conditions, being thus different for static sensors in a laboratory, transportable sensors in outdoor environment, and mobile sensors for onboard measurements. They rely on the use of isolation platforms, and auxiliary sensors like seismometers or accelerometers, eventually combined together, or on the development of better immune interferometer schemes~\cite{Fang2016}.

\section{\label{sec:gravity_sensors} Gravity sensors }

%\subsection{Introduction}
Gravity sensors are without any doubt the most emblematic inertial sensors based on atom interferometry. This stems from their relatively simple interferometer configuration, being a single axis vertical accelerometer, from their heritage, being one of very first demonstration of inertial sensing based on atom interferometry, and from the concrete application fields they address, in particular the field of geosciences.

\subsection{Historical context}
In their seminal paper \cite{Kasevich1991}, Kasevich and Chu used cold sodium atoms in an atomic fountain to realize the first demonstration of the 3-pulses atom interferometer based on Raman transitions, such as described in \ref{sec:principle}. In this early demonstration, the authors anticipated that such sensors could compete with state of the art classical gravimeters, such as based on the precise tracking of a free falling corner cube by optical interferometry.

Remarkably, in the following years, the efforts of Chu\ 's team made this claim become reality. A.~Peters \textit{et al} performed a very comprehensive metrological study of the performances, both in terms of stability and accuracy, of a second generation instrument based on Cs atoms \cite{ peters_measurement_1999, peters_high-precision_2001}. The stability reached a level as low as $20\times10^{-8}$~m.s$^{-2}$ at 1.3~s measurement time, and a large number of systematic effects were studied, and evaluated with a combined uncertainty of 3.2$\times10^{-8}$~m.s$^{-2}$. A direct side by side comparison with a commercial corner cube gravimeter, FG5 from the Microg solutions company \cite{Niebauer_1995}, showed a 4 times better stability for the atom gravimeter, and agreement between the two determinations of $g$, within the combined uncertainty of 7$\times10^{-8}$~m.s$^{-2}$.  At the same time, pioneering works on atomic gradiometers, which are differential accelerometers, have been conducted in Stanford \cite{ snadden_measurement_1998}, and a few years later at LENS \cite{bertoldi_atom_2006}.

% ***** FOR AVS Quantum science version *******************
%\begin{figure*}[t]
%	\centering
%	\includegraphics[width=\linewidth]{Gravis2.pdf}
%  	\caption{\textbf{Cold atom gravimeters} a) Scheme of the Cs based Stanford setup, with all critical functionalities represented. (Reprinted from A. Peters \textit{et al.}, "High-precision gravity measurement using atom interferometry", Metrologia \textbf{38}, 25-61 (2001) \copyright BIPM. Reproduced by permission of IOP Publishing. All rights reserved (Ref.~\cite{peters_high-precision_2001})); b) picture of the HUB gravimeter (Reprinted from M. Hauth Ph.D. Thesis " A  mobile,  high-precision  atom-interferometer  and its application to gravity observations, Humboldt-Universitat z\"u Berlin (2015) (Ref.~\cite{HauthPhD})), using the same fountain configuration as in a) but with Rb atoms; c) picture of the SYRTE gravimeter, with its magnetic shields opened, which simply drops Rb atoms, (Adapted from A. Louchet-Chauvet \textit{et al.}, "The influence of transverse motion within an atomic gravimeter", New J. Phys. \textbf{13}, 065025 (2011) (Ref.\cite{LouchetChauvet2011})) d) scheme of a single beam gravimeter, (Adapted from Q. Bodart \textit{et al.}, "A cold atom pyramidal gravimeter with a single laser beam", Appl. Phys. Lett. \textbf{96}, 134101 (2011), with the permission of AIP Publishing (Ref.\cite{bodart_cold_2010})).}  
% 	\label{fig:CAG}
%\end{figure*}
% *******************************************************************

% ************* for arxiv version ***********************************

\begin{figure*}[t]
	\centering
	\includegraphics[width=\linewidth]{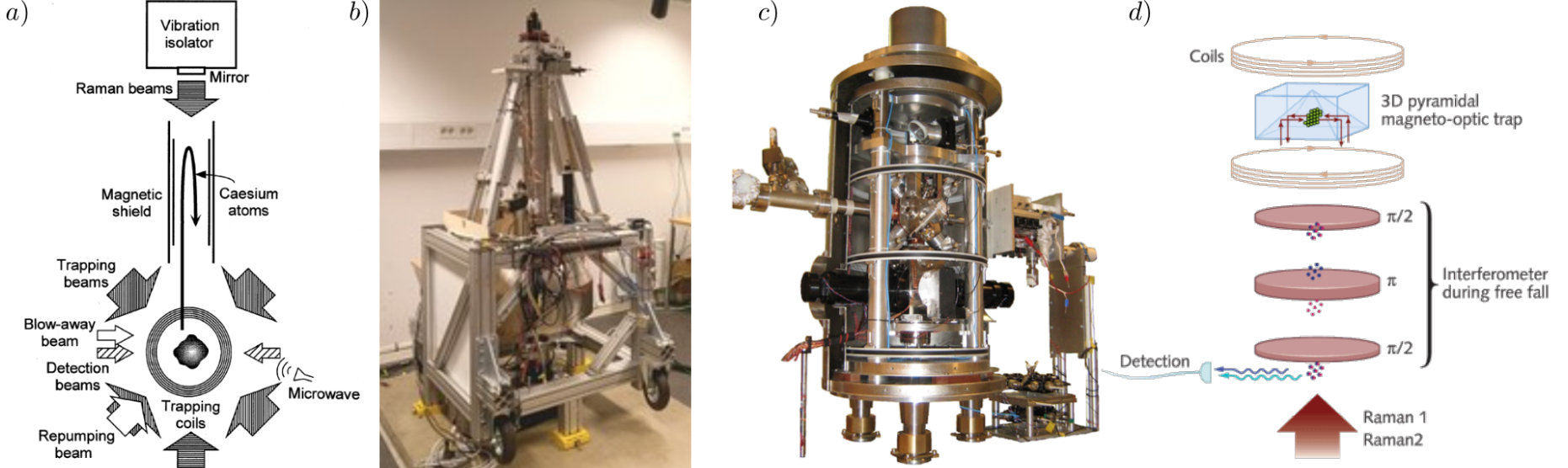}
  	\caption{\textbf{Cold atom gravimeters.} a) Scheme of the Cs based Stanford setup, with all critical functionalities represented. (Reprinted from A. Peters \textit{et al.}, "High-precision gravity measurement using atom interferometry", Metrologia \textbf{38}, 25-61 (2001) \copyright BIPM. Reproduced by permission of IOP Publishing. All rights reserved (Ref.~\cite{peters_high-precision_2001})); b) picture of the HUB gravimeter (Reprinted from Humboldt-Universit\"at zur Berlin website (Ref.~\footnote{HUB website: \url{https://www.physics.hu-berlin.de/en/qom/research}})), using the same fountain configuration as in a) but with Rb atoms; c) picture of the SYRTE gravimeter, with its magnetic shields opened, which simply drops Rb atoms, (Adapted from A. Louchet-Chauvet \textit{et al.}, "The influence of transverse motion within an atomic gravimeter", New J. Phys. \textbf{13}, 065025 (2011) (Ref.\cite{LouchetChauvet2011})) d) scheme of a single beam gravimeter, (Adapted from Q. Bodart \textit{et al.}, "A cold atom pyramidal gravimeter with a single laser beam", Appl. Phys. Lett. \textbf{96}, 134101 (2011), with the permission of AIP Publishing (Ref.\cite{bodart_cold_2010})).}  
 	\label{fig:CAG}
\end{figure*}
%*************************************************

Early works on gravimeters have triggered a wealth of developments. Projects aiming at more compact and transportable cold atom gravimeters started in the beginning of the 2000’s, in particular at SYRTE \cite{LouchetChauvet2011} and HUB \cite{ schmidt_mobile_2011}, and later by the AOSense company in the USA. Since then, the technology of atomic gravity sensors has considerably grown in maturity, as assessed by some major achievements, such as $i)$ the participation since 2009 to CIPM Key Comparisons (KC) and Euramet comparisons of absolute gravimeters\cite{jiang_8th_2012, francis_european_2013, francis2015}, in 2017, even if not included in the $3^{rd}$ KC\cite{Wu2020}, four atom gravimeters developed in China\cite{Wang2018, xu_-site_2018, Zhang2018, Fu2019} have participated to the associated pilot study; $ii)$ the demonstration of on board measurements, in a ship \cite{ bidel2018} and a plane \cite{bidel2019absolute} and $iii)$ the industrial development and commercial product offer of atom gravimeters at a competitive level of performance \cite{menoret_gravity_2018}. In total, about 30 research groups and private companies are today working on the development of atomic gravity sensors.

\subsection{Principle of the gravity measurement}

In a gravimeter configuration (Fig.\ref{fig:principle}.c), the sensitivity to rotation, the second term of  equation~\ref{eq:phase_shift_accel_rot}, is null as the two interferometer arms do not enclose any physical area. Raman transitions being velocity selective, one needs to chirp the frequency difference between the Raman lasers in order to compensate for the linear increase of the Doppler shift with time, and keep the three pulses on resonance. A frequency chirp $\delta \omega = at$ leads to an additional contribution in the interferometer phase given by $aT^2$, which allows for scanning the interferometer phase in a perfectly controlled way and record a fringe pattern. This chirp induced phase compensates the gravity phase shift when it perfectly matches the Doppler shift: $a_0 = \keff g$. This leads to a dark fringe in the interferometer pattern, whose position (as a function of the chirp rate $a$) does not depend on $T$. Indeed, in the reference frame of the free falling atoms, the phase difference between the Raman lasers is fixed, without any acceleration. Precisely locating, and tracking, this fringe allows for measuring $g$ via the relation $g=a_0/\keff$, as well as its fluctuations, in terms of SI traceable frequency measurements. This gives to this kind of sensors their absolute character: they do not require calibration, but an accurate control and knowledge of the RF and laser frequencies involved in the measurement. Remarkably, the cold atoms themselves, being well shielded from environmental perturbations in the drop chamber, can be used to guarantee the control and knowledge of these frequencies \cite{ xu_-site_2018}.

\subsection{Sensitivity limits}\label{sec:senslim}

\begin{figure}[t]
	\centering
	\includegraphics[width=\linewidth]{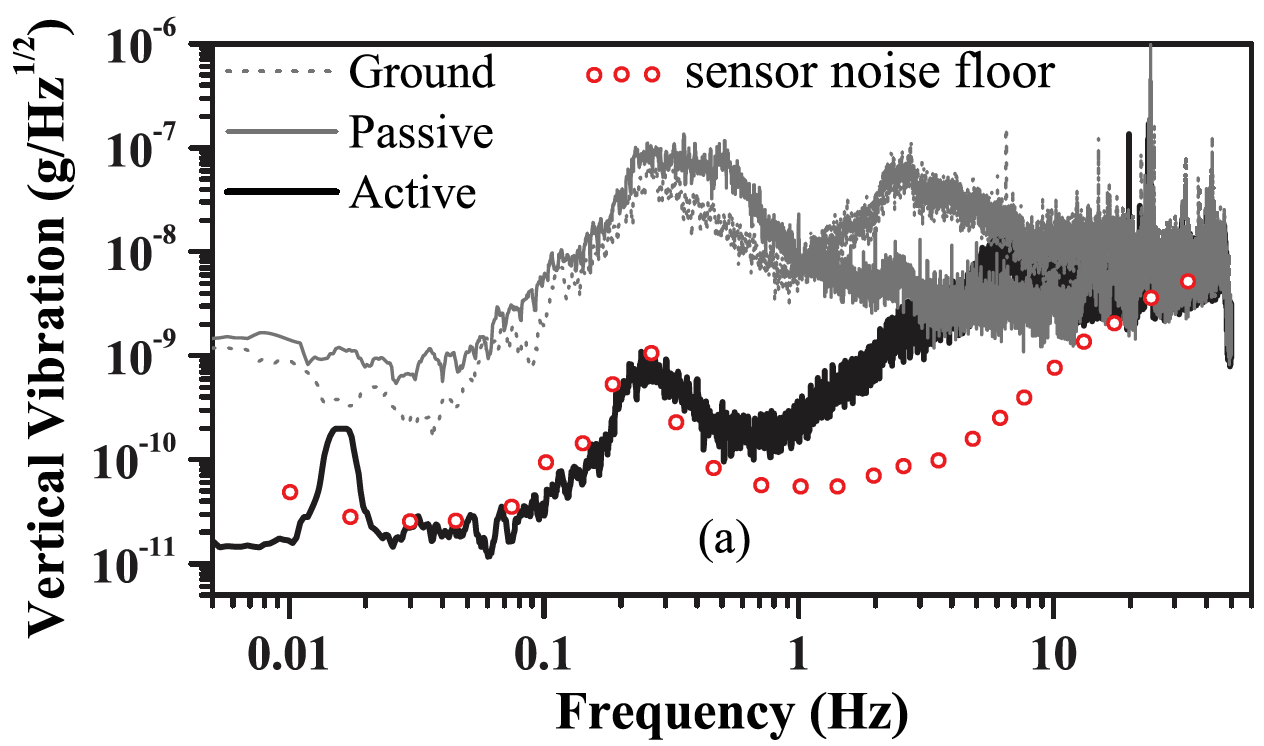}
	\caption{Amplitude spectral densities of vibration noise measured on the ground and on isolation platforms. Active isolation, based on sensing and actuation, allows to reach a level of vibration noise as low as the intrinsic noise of the sensor used for sensing. (Reprinted figure with permission from M.-K. Zhou \textit{et al.}, Phys. Rev. A \textbf{86} 043630 (2012). \copyright 2020 by the American Physical Society ( Ref.\cite{zhou_performance_2012})).}
	\label{fig:isolation}
\end{figure}

Typical interrogation times are in the range $2T~=~100$~ms to 1~s, depending on the size of the drop chamber (from about 10~cm up to 10~m). As mentioned in part~\ref{sec:noise} for increased $T$, parasitic vibrations limit the sensitivity. Different isolation methods have been used, based on superspring stabilization \cite{hensley_active_1999}, or the use of passive isolation platforms \cite{LeGouet2008}, eventually combined with additional active stabilization feedback control \cite{ zhou_performance_2012,tang_programmable_2014, zhou_note:_2015,Dong-Yun-2018}, or the correlation of the interferometer phase with the remaining vibration noise measured by a classical sensor, either a seismometer \cite{Merlet2009}, or an accelerometer \cite{lautier_hybridizing_2014}. The latter method allows for correcting the interferometer phase, either via postcorrection \cite{LeGouet2008}, or feed forward compensation in real time on the Raman lasers phase difference \cite{lautier_hybridizing_2014}. The latter scheme allows for operation without isolation platform \cite{Merlet2009}, and for efficient hybridization of classical and atomic accelerometers \cite{lautier_hybridizing_2014}. Figure \ref{fig:isolation} displays the amplitude spectral densities of vertical vibration noise measured on the ground and on such dedicated isolation platforms\cite{zhou_performance_2012}. These methods enabled several teams to reach sensitivities below 10$\times10^{-8}$~m.s$^{-2}.\tau^{-1/2}$ (Refs.~\cite{muller_atom-interferometry_2008, Hu2013, gillot_stability_2014, freier_mobile_2016}), even without isolation platform in quiet environment\cite{Farah2014}, the record being held by HUST\cite{Hu2013}, with $4.2\times10^{-8}$~m.s$^{-2}.\tau^{-1/2}$.

Many other sources of noise impact the gravity measurements, such as related to the phase noise of the RF reference frequency oscillators, frequency, phase and intensity noise of the Raman lasers, and detection noise. Detailed analysis of these noise sources have been carried out, especially in Ref.\cite{peters_high-precision_2001}, and later in Ref.\cite{LeGouet2008}, where the exact transfer function of the interferometer to phase noise fluctuations, the so-called sensitivity function derived in Ref. \cite{cheinet_measurement_2008}, was extensively used (with an extension to arbitrary pulse shapes presented in Ref.\cite{Fang2018}). These sources of noises can be reduced down to the mrad per shot level, well below the level of residual vibration noise, which still amounts to 10-100 mrad per shot, even with sophisticated vibration isolation schemes. 

The situation is different for gravity gradiometers, for which vibrations are a common source of noise, which is thus rejected in the differential measurement \cite{ snadden_measurement_1998}. This allows in principle these sensors to approach their intrinsic limit which will be set by detection noise, and ultimately by quantum projection noise. 

Best differential acceleration sensitivities of $3-4\times 10^{-8}\ \text{m.s}^{-2}.\tau^{-1/2}$ have been demonstrated with standard Raman interferometers \cite{mcguirk2002,sorrentino_sensitivity_2014}. For baselines of 1.4 m along the vertical direction \cite{mcguirk2002} and of 0.72 m along the horizontal direction \cite{biedermann_testing_2015}, this corresponds to best demonstrated gravity gradient sensitivities of $28$ and $59\times10^{-9}~\text{s}^{-2}.\tau^{-1/2}$.

\subsection{Accuracy limits}\label{sectionAccuracy}
\begin{figure}[t]
	\centering
	\includegraphics[width=0.8\linewidth]{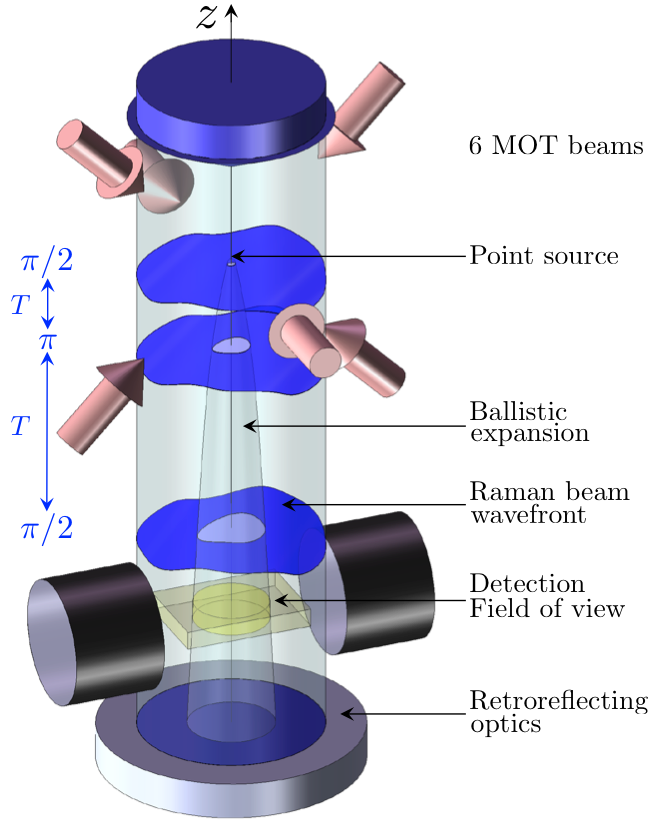}
	\caption{\textbf{Wavefront aberration effect.} (Adapted from R. Karcher \textit{et al.}, "Improving the accuracy of atom interferometers with ultracold sources", New J. Phys. \textbf{20}, 113041 (2018) (Ref.~\cite{karcher_improving_2018})). Because of to their ballistic expansion across the Raman beam, the atoms sample at the three $\pi/2-\pi-\pi/2$ Raman pulses different parasitic phase shifts related to wavefront distortions (displayed in blue as a distorted surface). This leads to a bias in the gravity measurement, resulting from the average of the effect over all atom trajectories, filtered by finite size effects, such as related to the waist and clear aperture of the Raman beam and to the finite field of view at the detection.}
  	\label{fig:ab}
\end{figure}

High accuracy is another appealing feature of atomic gravity sensors. Their scale factor being tied to time and frequency is well defined, which brings intrinsic accuracy and long term stability to the sensors. Yet, a number of systematic effects do bias the measurement, which have to be well measured and/or modelled, in order to correct the gravity measurements. A first detailed analysis, though not completely exhaustive, of systematic effects was carried out in Ref.\cite{ peters_high-precision_2001}. In principle, the interferometer is insensitive to frequency detunings of the Raman condition, but not to its gradients, leading to sensitivity to inhomogeneities of light shifts and magnetic field gradients. Remarkably, these shifts, as well as others related to electronic phase delays, can be efficiently eliminated via the so called $k_\text{eff}$ reversal technique \cite{weiss_precision_1994}. Indeed, the interferometer can be realized with two different orientations of the $k_\text{eff}$ wavevector (which corresponds to diffracting the intial wave packet upward or downward at the first beamsplitting pulse). This reverses the sign of the gravity phase shift, but not of the above mentionned systematic effects, so that averaging the $g$ measurements over the two directions cancels them. 
Some other effect do remain, the most important being a second order light shift \cite{ gauguet_off-resonant_2008}, the Coriolis acceleration \cite{peters_high-precision_2001} and the effect of laser wavefront aberrations \cite{LouchetChauvet2011} represented in figure~\ref{fig:ab}. 

The first effect is related to the presence of the second pair of Raman lasers, which is frequency detuned by twice the Doppler shift. This is a drawback associated to the use of retro-reflected Raman lasers, which allows on the other hand for reducing the impact of wavefront distortions, as detailed later. This results in a bias that is larger when dropping the atoms than when launching them upwards, and that scales with the Raman coupling. It can thus efficiently be corrected by performing $g$ measurements at different Raman intensities \cite{ gauguet_off-resonant_2008}. 

The second is related to residual transverse velocities, which lead to Coriolis accelerations. At $\mu$K temperatures, atoms have residual velocities of order of cm.s$^{-1}$, which leads to Coriolis accelerations as large as $10^{-6}$~m.s$^{-2}$. Hopefully, the net effect results from the average over the transverse velocity distribution. This leads in principle to a cancellation of the effect, provided that this velocity distribution is symmetric and well centered around zero, and that atoms with different transverse velocities do perform the interferometer and are detected with the same efficiency, or at last symmetrically with respect to the center of the velocity distribution. In practice, residual asymmetries and geometrical effects, such as due to laser beams inhomogeneities and finite size of the detection zone, can compromise this cancellation \cite{gillot_limits_2016}. The effect can then be evaluated by performing gravity measurements with different orientations by rotating the experiment \cite{LouchetChauvet2011}, the average between two opposite orientations allowing to cancel the effect. An efficient alternative consists in counter rotating the experiment \cite{ peters_high-precision_2001} or more simply the retroreflecting Raman mirror \cite{lan_influence_2012, Hauth2013} in order to compensate for the Earth rotation and thus eliminate the phase shifts induced by Coriolis acceleration. 

The third effect arises from the deviation of the Raman lasers wavefronts with respect to the ideally flat equiphase surfaces that act as a reference ruler used to measure the position of the free falling atoms at each of the three laser pulses. The required level of flatness is extremely demanding as a distortion as small as 0.1 nm corresponds to a Raman phase difference of about 1 mrad. The net effect results from the averaging of all possible trajectories, which samples differently the phase defects at each pulse. One can calculate that for the simple case of a curvature, and for an atomic temperature of 2 $\mu$K, a wavefront flatness of $\lambda/300$ PV over 1~cm is required to keep the error on the gravity measurement below $10^{-8}$~m.s$^{-2}$ (Ref.~\cite{karcher_improving_2018}). This implies that already the Gaussian character of the Raman laser beams has an effect, so that large size beam of waists larger than 1 cm are required to reduce the effect of the residual curvature. Finally, this effect is the most important contribution in the accuracy budget of the most accurate gravimeters, with contributions, up to recently, of order of $3-4\times10^{-8}$~m.s$^{-2}$. To evaluate this effect, one can investigate its dependence when increasing the atom temperature \cite{LouchetChauvet2011}, or when selecting the trajectories, either by truncating the detection area \cite{ schkolnik_effect_2015} or the size of the Raman laser beam \cite{ zhou_observing_2016}. But, none of these methods allowed to evaluate the effect with a low enough uncertainty to make the accuracy of the atomic sensor better than the announced one of the best classical instrument, the FG5 (Ref.~\cite{Niebauer_1995}) or FG5-X (Ref.~\cite{Niebauer_2012}) corner cube gravimeters. A better method consists in extrapolating the bias to zero temperatures by performing $g$ measurements as a function of decreasing temperatures. Such measurements, using ultracold atoms produced via evaporative cooling in a dipole trap, recently allowed for extracting a model of the wavefront and reducing the uncertainty of the wavefront aberration bias down to $1.3\times10^{-8}$~m.s$^{-2}$ (Ref.~\cite{ karcher_improving_2018}).

%\begin{figure}[t]
%	\centering
%	\includegraphics[width=\linewidth]{3D-Aberration.png}
%	\caption{\textbf{Aberrations of a window. } }
%  	\label{fig:abwindow}
%\end{figure}

\subsection{Commercial instruments}

Even though some of the best atomic instruments have demonstrated performances better than state of the art classical sensors, many of these devices are "laboratory sensors" in the sense that they mostly operate in laboratory conditions, with air conditioning system for example. This is enough for some applications, but not for on field measurements, for which instruments have to be robust, compact and easy to operate for non-physicist operators, and sustain large temperature and humidity changes. Some companies have embarked on this path. AOSense was formed in 2004 to spin-off innovative research developed at Stanford University and delivered its first commercial compact gravimeter to an aerospace customer in 2010 (Ref.~\footnote{AOSense website: \url{https://aosense.com}}). Since then, other companies, mainly in Europe have followed. They are listed in the tables presented in the section \ref{sec:list_groups}. One of them is the Muquans\footnote{muquans website: \url{https://www.muquans.com}} company founded in 2011. Their products are the result of a long-term research effort initiated by SYRTE and LP2N. They have been developing commercial gravity sensors based on the simple architecture demonstrated in Ref.\cite{bodart_cold_2010} and on the ease of use and robustness of fibered laser systems\cite{stern_light-pulse_2009}. First gravimeters have already been delivered to customers from the  geophysics community\cite{menoret_gravity_2018}.

\begin{figure*}[t]
	\centering
	\includegraphics[width=0.7\linewidth]{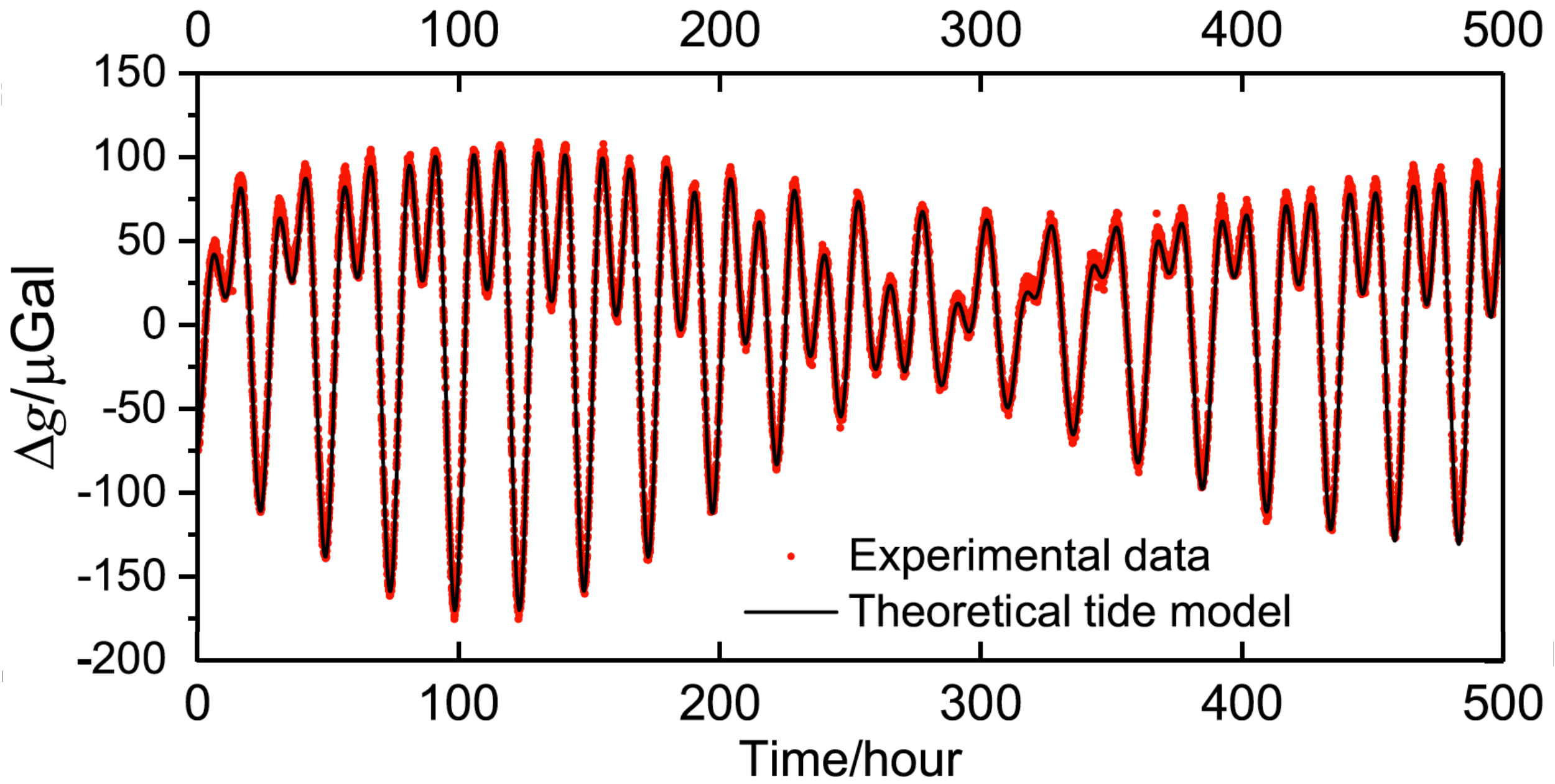}
	\caption{\textbf{Tide signals.} Example of NIM-AGRb-1 continuous gravity measurement over 20 days expressed in $\mu$Gal (1~$\mu$Gal~=~$10^{-8}$~m.s$^{-2}$). Each data point is a 3 min average. The quantum sensor sensitivity allows for measuring gravity changes due to tides, air pressure variations, polar motion, water table level fluctuations. (Adapted from S.-K. Wang \textit{et al.}, "Shift evaluation of the atomic gravimeter NIM-AGRb-1 and its comparison with FG5X", Metrologia \textbf{55}, 360-365 (2001) \copyright BIPM. Reproduced by permission of IOP Publishing. All rights reserved (Ref.~\cite{Wang2018})).
	}
	\label{fig:tides}
\end{figure*}

\subsection{Applications} 
Gravity sensors find applications in many fields, ranging from geophysics and geodesy, navigation, civil engineering and fundamental physics. So far, atom gravimeters have been mainly developed in or for laboratory-type environments, where they can reach excellent short term and long term stability, better than classical corner cube gravimeters \cite{Hu2013, gillot_stability_2014,freier_mobile_2016}. There, they allow for recording continuous series \cite{LouchetChauvet2011, freier_mobile_2016, Wang2018}, a mode of operation usually restricted to relative, spring or superconducting, gravimeters. Figure~\ref{fig:tides} displays an example for such signal. Being accurate, they can be used as metrological standards in National Metrology Institutes, and thus participate to CIPM Key Comparisons \cite{ jiang_8th_2012, francis2015}. Other applications in the field of metrology are $i)$ the precise determination of $g$ for Kibble balance experiments \cite{ thomas_first_2015, thomas_determination_2017}, which are now primary standards that realize the kilogram based on its new definition linked to the Planck constant \cite{Schlamminger2018, Stock2019}, and $ii)$ the determination of the gravitational constant $G$ at the $10^{-4}$ level with gravity gradiometers \cite{ fixler_atom_2007, rosi_precision_2014}.

\begin{figure}[t]
	\centering
	\includegraphics[width=0.7\linewidth]{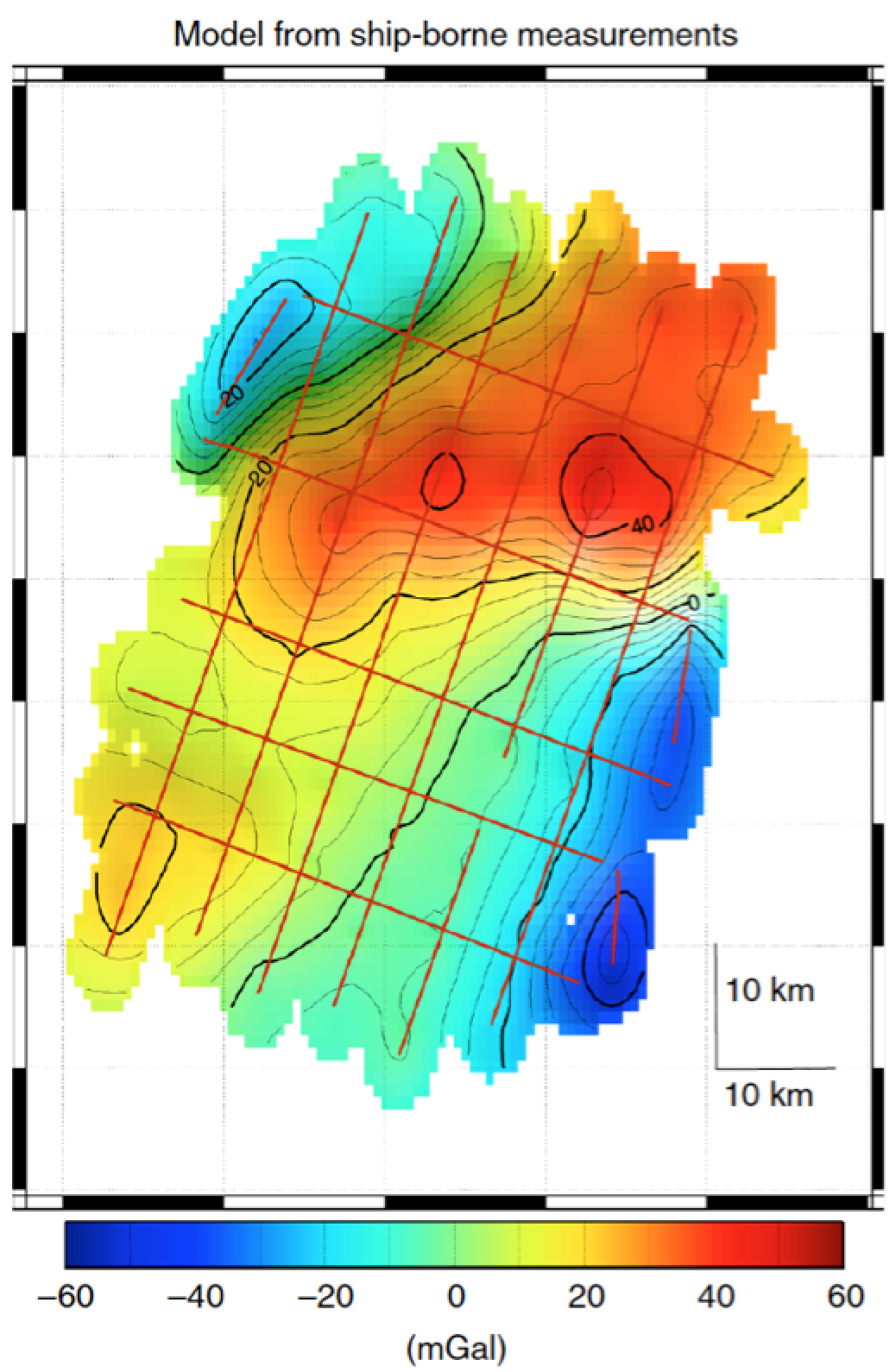}
	\caption{\textbf{Gravity anomaly model of Meriadzec terrace}. (Adapted from Y. Bidel \textit{et al.}, "Absolute marine gravimetry with matter-wave interferometry", Nat Commun \textbf{9}, 627 (2018) (Ref.~\cite{bidel2018})). Gravity is expressed in mGal (1~mGal~=~$10^{-5}$~m.s$^{-2}$). The model was obtained from ONERA ship-borne atom gravimeter measurements.}
  	\label{fig:gravmodelONERA}
\end{figure}

To address wider applications, engineering and simplification efforts in the sensors or its key subsystems are made in order to allow performing measurements in more aggressive environments \cite{ cheinet_compact_2006, bodart_cold_2010, schmidt_portable_2011, geiger_detecting_2011, bidel_compact_2013, menoret_dual-wavelength_2011, de_angelis_isense:_2011, schkolnik_compact_2016, Zhang2018, menoret_gravity_2018, Caldani2019}. Operation of gravity sensors on board a ship \cite{bidel2018}, and more recently in an aircraft \cite{bidel2019absolute}, has been demonstrated, and gravity mapping have been performed in both cases, showing improved repeatability and lower uncertainties compared to classical marine gravimeters. Figure~\ref{fig:gravmodelONERA} displays such gravity mapping. Ongoing developments target the deployment of gravity sensors, such as gradiometers, for civil engineering applications \cite{ boddice_capability_2017} or the installation of a commercial atom gravimeter for natural risk management, e.g. on the Etna volcano \footnote{Newton g project: http://www.newton-g.eu/index.htm}.

\section{\label{sec:inertial_sensors} Inertial sensors }
%\textcolor{red}{\textit{Accelerometers and gyroscopes developed by the different teams. Both compact and non compact. Present typical applications (navigation, geophysics) and remaining challenges.}}

\subsection{Gyroscopes}
\label{subsec:gyro}
Sensing rotations  with an atom interferometer belongs to one of the pioneering experiments from 1991 which triggered the field of atom interferometry \cite{Riehle1991}. In that study, a calcium atomic beam was excited in an optical Ramsey geometry \cite{Borde1989} on the intercombination line $^1S_0\rightarrow ^3P_1$ ($\lambda=657.46$~nm). The whole atomic-beam apparatus was mounted on a rotational stage and could be rotated around a vertical axis perpendicular to the plane defined by the laser beams and the atomic beam and the authors observed a phase shift proportional to the rotation frequency of the apparatus. Differences between rotation rates  of the order of 0.1~\radps could be resolved by the apparatus.

Following this proof-of-principle experiment, few groups developed interferometers using atomic beams as rotation sensors. The most important achievements were from the Pritchard group at MIT in 1997 \cite{Lenef1997} using nano-fabricated gratings to realize beam-splitters and mirrors, and from the Kasevich group at Stanford \cite{Gustavson1997,Durfee2006}  and Yale \cite{Gustavson2000}  using stimulated two-photon Raman transitions as atom-optics. A review on the historical aspects of these developments is presented in Ref.~\cite{Barrett2014}, which compares the performances of atomic beam and cold-atom based gyroscopes. While the short-term sensitivity of atomic beam interferometers still holds the record for atomic gyroscopes ($6\times 10^{-10}$~\radps at 1~s integration time) owing to the large atom flux \cite{Gustavson2000}, achieving long-term stability levels competitive with those of other technologies (e.g. optical gyroscopes) was challenging. On the contrary, the use of cold-atoms leads to a reduced atom flux, which limits the short term sensitivity, but allows a better control of atomic trajectories which is advantageous to achieve better long term stability levels. We will therefore focus here on cold-atom based gyroscopes.

\subsubsection{Rotation phase shift.}
As explained in section \ref{sec:principle}, a phase shift will appear in an interferometer where the atomic wavepacket moves with a velocity $\vec{v}$ with respect to a frame rotating at an angular velocity $\vec{\Omega}$, given by 
\begin{equation}
\Phi\stxt{rot} = \veckeff\cdot(2\vec{\Omega}\times \vec{v}) T^2,
\label{eq:phase_gyro}
\end{equation}
where  $T$ is the time between the light pulses (Mach-Zehnder geometry assumed here).
An argument which is often put forward to explain the potentially very large sensitivity of atomic gyroscopes compared to their optical counterparts is based on the expression of the phase shift of the Sagnac effect\cite{Sagnac1913}:
\begin{equation}
\Phi\stxt{rot}=\frac{4\pi E}{h c^2}\vec{\Omega}\cdot\vec{A}.
\label{eq:phase_sagnac}
\end{equation}
In this expression, which ties to the (special) relativistic nature of the Sagnac effect (pointed out by von Laue  \cite{Gourgoulhon2013}), $E$ is the total energy of the particle associated with the interfering waves and $\vec{A}$ is the oriented area-vector of the interferometer (the prefactor is $8\pi E$ in the case of a full loop interferometer). In the case of non-relativistic  atoms, $E\simeq mc^2$ is about 11 orders of magnitude larger than the energy $h\nu$ of a photon used in an optical gyroscope, yielding a much larger scale factor for atomic gyroscopes. This increase in scale factor has nevertheless to be confronted to the much larger photon flux and the larger area (e.g. in a fiber optic gyroscope) in  optical interferometers.
While this formulation helps in assessing the potential of an atomic gyroscope over an optical gyroscope, it can lead to controversies in interpretation on the actual origin of the phase shift (see, e.g. Ref.~\cite{Wolf2011}), which, as shown in section \ref{sec:principle} solely originates from the sampling of the laser wavefront by the motion of the atomic wavepacket in the laser frame.
The link between Eq.~\eqref{eq:phase_gyro} and Eq.~\eqref{eq:phase_sagnac} is  obtained by calculating the area of the interferometer given by 
\begin{equation}
\vec{A}=\frac{\hbar\veckeff}{m} T\times \vec{v}T.
\end{equation}
The larger scale factor of the atomic gyroscope over the optical gyroscope at constant interferometer area can  be understood from the fact that the atom travels (at velocity $v$) slowlier than the photon (velocity $c$) in the interferometer of fixed dimensions, thereby sensing the inertial effect for an increased duration.
%An alternative (rather more intuitive than rigorous) picture to assess the larger scale factor of the atomic gyro is to simply compare the time spent by the particle in an interferometer of given area. Indeed, the phase shift in an optical Mach-Zehnder interferometer can formaly be expressed in the same way as for an atom interferometer (Eq.~\eqref{eq:phase_gyro}) as $\Phi\stxt{rot}\STXT{phot}=k~ (2\Omega \times c) T^2$, where $T=L/c$ is the time spent by the photon is the arm of a  square-shaped interferometer of arm length $L$ and $k=2\pi\nu/c$ is the photon wavevector. The ratio of scale factors between the atom and photon gyroscopes there appears as $c/v$, which means that in an interferometer of fixed physical dimensions, the atom will spend much more time to sense the inertial effect than a photon travelling at $c\gg v$. 

This consideration on the importance of the interrogation time $T$ drives the design of atomic gyroscopes. We will present below two complementary directions of research: on one side some experiments target large interrogation times in order to increase the physical area, which supposes a large interrogation region since the free fall distance of the atoms scales as $T^2$; on the other side some efforts are conducted to reduce the physical dimensions of the sensor at the cost of a reduced sensitivity.

\subsubsection{Instruments targeting high stability levels with long interrogation times}

\paragraph{First generations of cold-atom gyroscopes.}

The first cold-atom gyroscope experiment was started at the SYRTE laboratory in 2000 and developed until 2008. It used two clouds of Cesium atoms launched along parabolic trajectories and traveling in opposite directions through a common interrogation region, where three pairs of retro-reflected Raman beams enable to realize a full inertial basis (the three components of rotation and the tree components of acceleration) \cite{Fils2005,Canuel2006}. In particular, the use of two counter-propagating atom clouds enabled to separate the horizontal acceleration and vertical rotation components of the phase shift. Finally, the demonstration of principle of  a four pulse interferometer sequence allowing to perform the measurement of one component of rotation without DC acceleration sensitivity was  demonstrated.  Ref.~\cite{Gauguet2008} presents the complete characterization of the instrument when measuring one horizontal rotation axis: with a total interrogation time $2T=80$~ms, the authors demonstrated a sensitivity of $2.4\times 10^{-7}$~\radps.$\tau^{-1/2}$ limited by quantum projection noise, a long term stability of $1\times 10^{-8}$~\radps at 4000~s integration time and a test of the linearity of the scaling factor at one part per 10$^5$. The limitations to the stability were identified as coming from the fluctuations of atom trajectories coupled to the wave-front distortions of the Raman laser.

\begin{figure*}[t]
	\centering
	\includegraphics[width=0.8\linewidth]{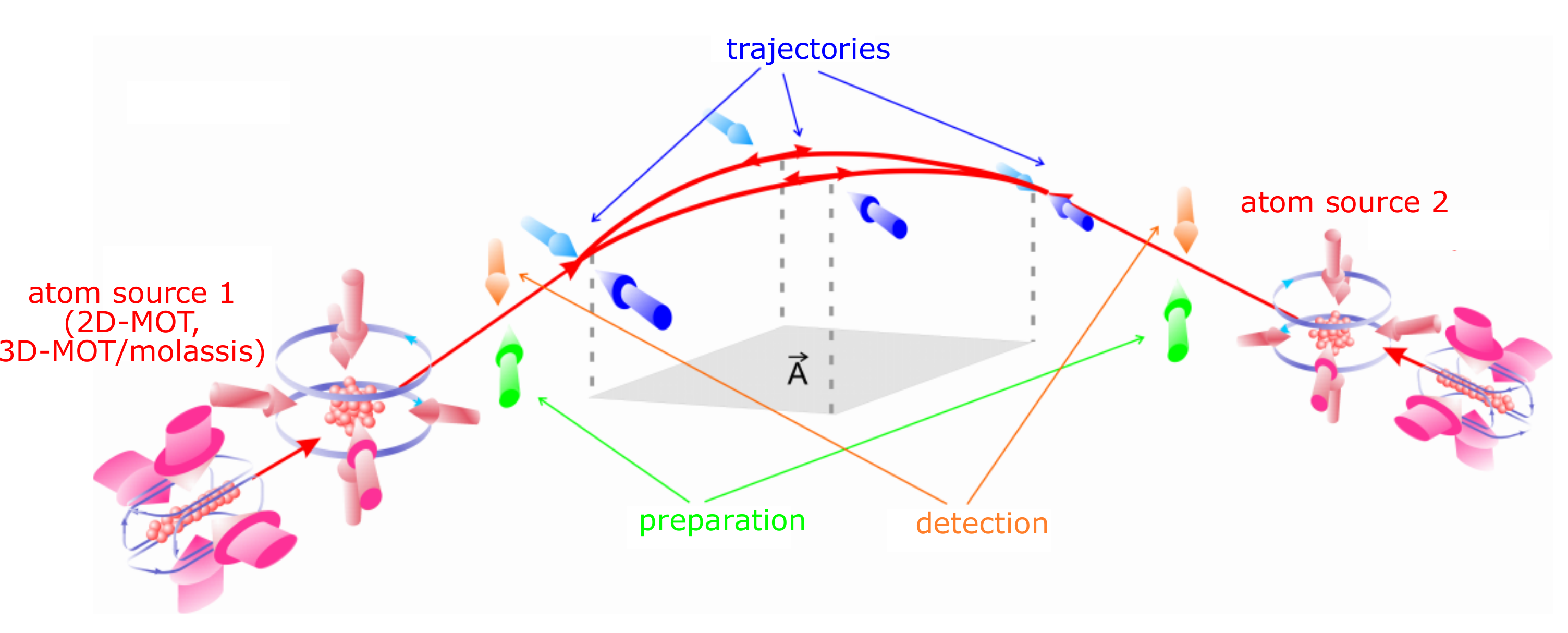}
	\caption{Cold-atom gyroscope with counter-propagating cold-atom sources developed at the University of Hannover. (Figure adapted from Ref.~\cite{jentsch2004konzeption}).}
	\label{fig:rasel_gyro}
\end{figure*}

About at the same time, a cold Cs-atom gyroscope was developed at Stanford University. The experiment volume was comparable to the SYRTE gyroscope but used an architecture dedicated to the four pulse sequence: two clouds of atoms were also prepared in two different regions but launched along a vertical trajectory as in an atomic fountain, and interrogated by four-light pulses ($\pi/2-\pi-\pi-\pi/2$, total interrogation time of 206~ms). The details of the apparatus developed during the period from 2002 to 2010 are given in Ref.~\cite{TakasePhD2008} with the end results published in 2011 in Ref.~\cite{Stockton2011}. 
In this experiment, the authors demonstrate how this four pulse configuration overcomes accuracy and dynamic range limitations of three pulse  atom interferometer gyroscopes. Moreover, by introducing a time asymmetry in the sequence, they present a method to suppress spurious noise terms
related to multiple-path interferences, thereby increasing the signal to noise ratio of the interferometer. They show how the instrument can be used for precise determination of latitude, azimuth (true north), and Earth’s rotation rate, and highlight the large potential of the four pulse configuration. 
%In particular,  a rotation rate sensitivity of $8.5\times 10^{-8}$ \radps \ was obtained by correcting in real time the inertial noise\textcolor{red} {il faut en dicuter car il ne demontre pas une capacité de mesurer la rotation à ce niveau, c'est uniquement le RSB. si je me rappelle bien ils montrent une corrélation mais le chiffre lui n'est pas issu de ce traitement }. The scale factor of the gyroscope was also studied, both for the Earth rotation rate and for rotations with respect to the local geostationary reference frame.\textcolor{red}{à rediscuter tous les deux également Remi } 

Few other cold-atom gyroscope projects have been conducted. At the University of Hannover (Germany), a  sensor of 13.7~cm length  was developed, featuring a Sagnac area of 19~mm$^2$ (Fig.~\ref{fig:rasel_gyro} and Ref.~\cite{Mueller2009_rasel}). This experiment used a three light pulse ($\pi/2-\pi-\pi/2$) configuration with two clouds of atoms launched horizontally in opposite directions from two sources, and crossing three physically separated interaction regions. In particular, a method to minimize the relative alignment of the three Raman beams was demonstrated in Ref.~\cite{Tackmann2012} by maximizing the contrast of the interferometer. A short term sensitivity of $6.1\times 10^{-7}$ \radps.$\tau^{-1/2}$ was achieved. A modified version of this setup to accommodate composite light pulses that mitigate some of the technical noise sources (e.g. light shifts) was reported in Ref.~\cite{Berg2015}, where a sensitivity of $1.2\times 10^{-7}$ \radps.$\tau^{-1/2}$ was achieved at short time-scales (below 10~s). Another experiment is currently being developed in China \cite{Yao2018} on a similar basis, i.e. with separated atomic sources launched along parabolic trajectories and a total interrogation time of $104$~ms. In that setup, the current long term stability level reaches $6.2\times 10^{-8}$ \radps after 2000~s of integration time.

\paragraph{Strategy for improved long-term stability.}

As identified in Refs.~\cite{Fils2005,Gauguet2008}, the main limitation to the long-term stability of a cold-atom gyroscope is linked with the fluctuation of the atom' mean trajectory coupled to the imperfect relative wavefront of the Raman beams. As discussed in section \ref{sec:gravity_sensors}, the atoms probe a finite region of the laser wavefront which originates in a bias on the inertial measurement if not perfectly flat. This effect is even more pronounced in gyroscopes based on physically separated interrogation beams (in contrast to a single retro-reflecting mirror in the case of the gravimeter). Fluctuations of the atomic trajectory then result in a random sampling of the Raman beam relative wavefront, yielding an instability. In most experiments, the effect of initial  velocity fluctuations $\delta v_0$ dominates that of initial position fluctuations. The source of fluctuations then  scales as $\delta\phi\stxt{wf}\simeq \frac{4\pi\delta v_0 T \delta \lambda}{\lambda^2}$, where $\delta \lambda$ is the deviation  of the wavefront with respect to a plane wave (for which $\delta \lambda =0$). Since the inertial signal scales at least with $T^2$ (see Eq.~\ref{eq:phase_gyro}), it is interesting to increase the interrogation time to minimize the bias with respect to the signal, at the cost of increasing the size of the instrument (which  scales with $v.T$ or $T^2$ as a function of the architecture). As a numerical example, achieving long-term stability levels  $\delta v_0<1 \ \mu$m.s$^{-1}$ is technically challenging, which results, for wavefront aberrations $\delta\lambda=\lambda/50$ over a typical pupil diameter of 1~cm, in an interferometer bias $\delta\phi\sim 15$~mrad  ($T=50$~ms). This bias has to be compared to a signal of about 16~rad for the Earth rotation rate ($72\ \mu$rad.s$^{-1}$) in an interferometer with $T=50$~ms and $v\simeq 3$~m.s$^{-1}$. The long term stability is thus constrained to a level of the order of $7\times 10^{-8}$~\radps, consistent with the value reported in several articles \cite{Gauguet2008,Berg2015,Yao2018}.

\begin{figure*}[t]
	\centering
	\includegraphics[width=\linewidth]{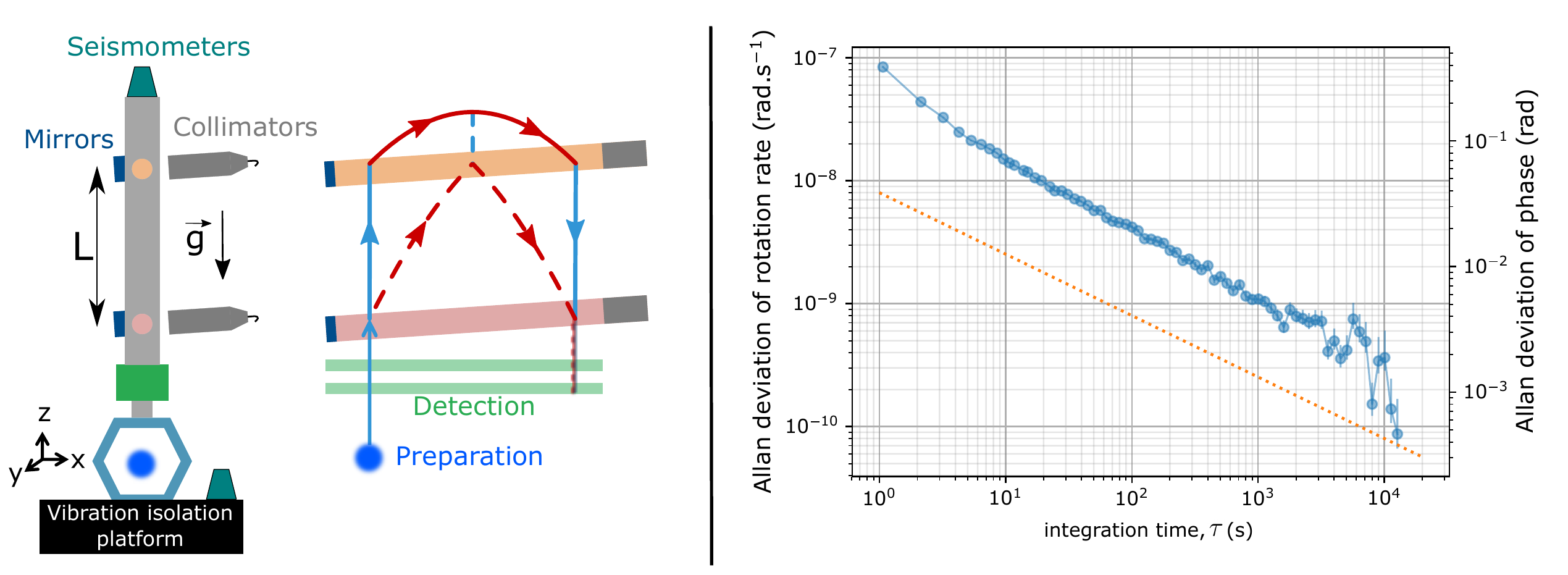}
	\caption{\textbf{Cold-atom gyroscope with a 11~cm$^2$ Sagnac area}. Left panel: sketch of the experiment presented in Ref.~\cite{Savoie2018}: a cloud of Cesium atoms is launched in the vertical direction and interrogated by four light-pulses; the middle panel shows the two arms of the interferometer. Right panel: the points show the rotation rate stability (Allan deviation) which is limited by vibration noise; the dotted line shows the limit associated with detection noise. (Adapted with permission from D. Savoie \textit{et al}, Science Advances, Vol. 4, no. 12, eaau7948 (2018) (Ref.~\cite{Savoie2018})).}
	\label{fig:SYRTE_4p_gyro}
\end{figure*}

\paragraph{Second generation of cold-atom gyroscopes.}

In that context, a new instrument was built at SYRTE to target better long term stability levels by increasing the interrogation time to $800$~ms. In the experiment described in Ref.~\cite{Dutta2016}, a four-light pulse architecture with an atom cloud launched vertically in a fountain configuration  as in \cite{Stockton2011}. In that setup, two beams separated by 58~cm perform the atom optics (Fig.~\ref{fig:SYRTE_4p_gyro}). In such a folded geometry, the interferometer phase shift acquires a cubic dependence with $T$ and is given by $\Phi\stxt{rot}=\frac{1}{2}\veckeff\cdot (\vec{g}\times \vec{\Omega})T^3$.  The phase shift associated with Earth rotation rate ($72\ \mu$rad.s$^{-1}$) becomes 333~rad (the Sagnac area is $11$~cm$^2$). Moreover, due to the folding of the trajectory, some effects of wavefront aberrations are reduced. This setup therefore currently represents the state-of-the-art for atomic gyroscopes, with a long term stability of $3\times 10^{-10}$ \radps after $10 \ 000$~s of integration time (Fig.~\ref{fig:SYRTE_4p_gyro}) \cite{Savoie2018}. 
Note that  the instrument still operates well above the quantum projection noise limit, which equals $2\times 10^{-10}$ \radps.$\tau^{-1/2}$ for $10^{6}$ atoms participating to the interferometer and a contrast of $50\%$ (assuming a cycle rate of $4$~Hz as in Ref.~\cite{Savoie2018}).
If the bias associated with the  imperfect atomic trajectories is controlled at a sufficient level (Ref.~\cite{Altorio2019}), then stability levels  in the range of   $10^{-12}$ \radps can be anticipated.

\subsubsection{Example of gyroscope simplification effort}
\begin{figure}[t!]
	\centering
	\includegraphics[width=\linewidth]{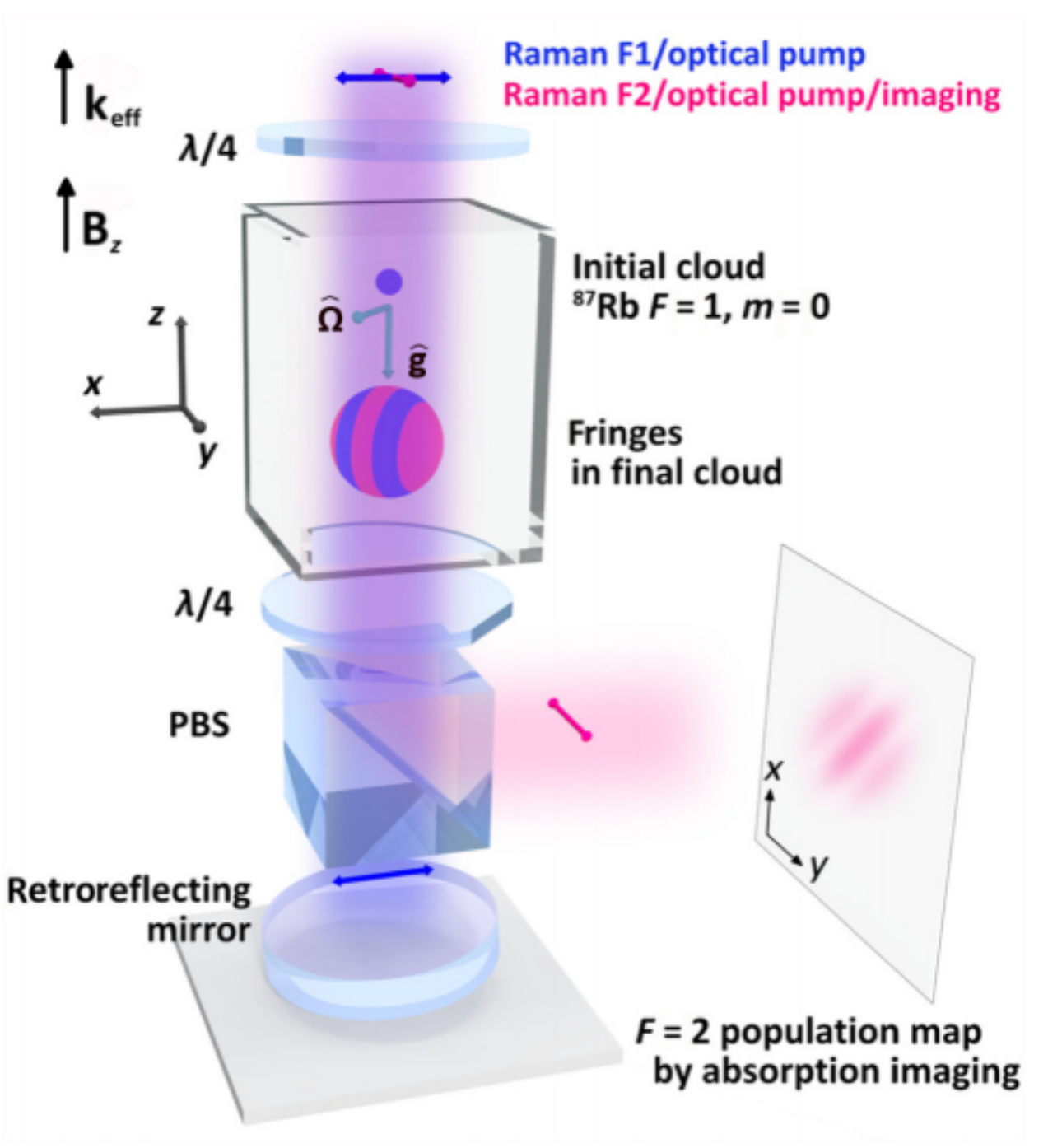}
	\caption{\textbf{Point-source interferometry in a centimeter-scale chamber}. 
	 $^{87}$Rb atoms are laser-cooled in a glass cell with $1$~cm$^2$ cross-section area. A manifold that includes an ion pump, a Rb dispenser, and a vacuum window is connected to the glass cell. In the glass cell, the laser beams for state preparation, Raman interrogation, and detection propagate vertically in a shared beam path with a $1/e^2$ beam diameter of 8~mm and are circularly polarized inside the glass cell. Three orthogonal and retroreflected beams (not shown in the figure) with $1/e^2$ diameters of 6~mm form the magneto-optical-trap.
	%(b) Allan deviation of rotation rate. The dashed black line is the performance of an ideal point source limited by atom shot noise. 
	The achieved short term sensitivity is currently $5\times 10^{-4}$~\radps.$\tau^{-1/2}$ and integrates to $5\times 10^{-5}$~\radps after 800~s of integration time. An improved version of the sensor could achieve a sensitivity of $10^{-6}$~\radps.$\tau^{-1/2}$. 
	(Reproduced with permission from Y.-J. Chen \textit{et al}, Phys. Rev. Applied 12, 014019 (2019). Copyright 2019, American Physical Society (Ref.~\cite{Chen2019})).
}
	\label{fig:PSI_NIST}
\end{figure}

In parallel to these developments aiming at achieving stability levels that could beat those from other navigation technologies in the future (see section below), efforts are conducted to study simplified architectures of sensors with a smaller volume. As an example, a group at NIST has built an instrument with a glass vacuum chamber occupying a surface of $1$~cm$^2$ (Fig.~\cite{PSI_NIST}) in which the rotation (and acceleration) phase shift can be measured by observing its dependence on the individual atom velocities according to Eq.~\eqref{eq:phase_gyro}. In this so-called point source interferometry (PSI) technique, originally demonstrated in a 10-meter long instrument \cite{Dickerson2013}, the interferometer configuration uses the natural expansion of the cold atom sample due to it's residual temperature. If the initial size of the atom cloud is negligible with respect to the size after expansion, a camera, which images the atom cloud, resolves in a position-dependent way the rotation phase shift (Fig.~\ref{fig:PSI_NIST}).   The Sagnac area is given by the root-mean-square atomic velocities and equals $0.03$~mm$^2$ in this setup. The authors identify the limitations to the sensitivity from the short Raman interrogation time ($T = 8$~ms), the technical noise, the initial size of the cold-atom cloud, and the measurement dead time. Moreover, they show how the instrument could be used for gyro-compassing.

While the PSI technique provides experimental simplicity, the scale factor of the sensor is dependent on the initial size of the atom cloud which is exaggerated in compact designs where the expansion ratio is small, which leads to  instabilities. In Ref.~\cite{Avinadav2020}, the authors show how using additional information on the contrast and cloud size from the PSI images allows to determine the scale factor correction in each image, thereby enabling to suppress scale factor drifts by a factor 10 without degrading the short term sensitivity.

Other efforts by several other teams are ongoing, in order to build complete inertial measurement units (i.e. 3 axes of acceleration and 3 axes of rotation) in a compact system. They will be presented in the section below discussing accelerometer developments \ref{subsec:accelero}.

\subsubsection{Applications}
We discuss here some applications of gyroscopes and relate them to different   technologies in order to appreciate the efforts that must be accomplished by the community to enlarge the potential of cold-atom sensors.
The appealing feature of cold-atom gyroscopes relies in their inherent long-term stability associated with the stability of the quantities involved in the gyroscope scale factor. This feature, associated with the complexity of instruments, naturally points for high-performance applications such as strategic inertial navigation, or scientific applications, e.g. in geosciences or tests of fundamental physics.

\paragraph{High-performance inertial measurement unit.}
As discussed above, the  limit to the stability of cold-atom gyroscopes currently lies between $10^{-9}$ and $10^{-10}$ \radps \ for the best instruments \cite{Savoie2018}.
Several other gyroscope technologies address the application of navigation. Current developments of MEMS gyroscopes target instability levels below $0.05~^\circ/h$ ($2.5\times 10^{-7}$ \radps) (Ref.~\cite{Cui2019}), which makes this technology  very important for military applications given their level of integration \footnote{Website: \url{https://issuu.com/globalbusinessmedia.org/docs/defence_industry_reports_95________}}.
This can be compared to the best strategic-grade fiber optics gyroscopes (FOG)  which feature instability levels in the  $10^{-10}$ \radps \ range (Refs.~\cite{Lefevre2014,blueseis,Korkishko26-2}), or to Hemispherical Resonator gyroscopes (HRG) with comparable performances \cite{Delhaye26-2}. 
To refine the comparison, it is worth mentioning the importance of additional properties of sensors such as dynamic range, number of measurement axes, robustness to the environment (vibrations, temperature fluctuations), level of integration and industrialization possibilities. A large research effort in these direction is required to enlarge the scope of applications of cold-atom gyroscopes. As a result, the cold-atom technology will probably, in a first stage, address applications requiring high stability levels but operating in quiet environments, for example in underwater navigation (e.g. in a submarine);

\paragraph{Scientific applications.}
Large ring laser gyroscopes (RLG, Ref.~\cite{Schreiber2013}) feature instability levels in the $10^{-14}$ \radps \ range, which offers possibilities to study the evolution of the Earth polar motion \cite{Schreiber2011}. Their excellent short term sensitivity is also exploited  for rotational seismology \cite{Hadziioannou2012}: here, colocalized acceleration (with seismometers) and rotation (with the RLG) measurements  can  inform on the direction of propagation of seismic waves as well as on the phase velocity of the waves, which represent a key information for geophysics. Cold-atom sensors, which can measure in a single platform and at the same position several components of the local instantaneous rotation and acceleration vectors could have a large impact in this emerging field. While their rotation rate sensitivity  still not competes with  that of RLG, this technology is interesting as it is in principle transportable, while  current RLG are rather infrastuctures than sensors. The accurate knowledge of the scale factor in combination with portability could allow to spatially distribute several sensors in order to perform correlative rotational seismology. Much progress is expected in this direction together with the improvement of the short term sensitivity levels of cold-atom gyroscopes.

Finally, cold-atom gyroscopes have been proposed for tests of fundamental physics. The most accomplished project has been a test of  general relativity by measuring the Lense-Thirring effect in space \cite{LANDRAGIN2002,Jentsch2004}, where both the quiet environment and long interrogation times are suitable for the very high sensitivity required in such a test.

\subsection{Accelerometers and progress for a complete inertial measurement unit}
\label{subsec:accelero}
Accelerometers using atom-interferometry are mostly developed for applications to inertial navigation as a building block of a full inertial measurement unit (IMU) \cite{Canuel2006,Battelier2016}. As a single-axis accelerometer only requires one laser beam interacting with atoms at one given position, their implementation is simpler than for gyroscopes, which need to open a physical area to the interferometer. However, the first demonstration in a mobile vehicle \cite{geiger_detecting_2011} (a plane) has shown the difficulties due to high frequency accelerations (vibrations) of the carrier and dead times between successive measurements,  since the use of an isolation platform is not a solution for this type of operation. Different approaches have been demonstrated or proposed to overcome these difficulties and will be detailed in the following.

\subsubsection{Increasing the repetition rate and bandwidth}
\begin{figure}[h]
	\centering
	\includegraphics[width=0.8\linewidth]{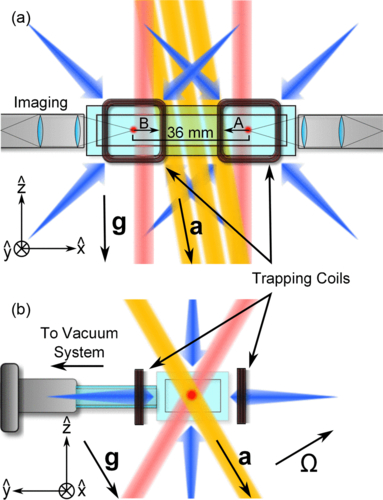}
	\caption{\textbf{Dual-axis high-data-rate atom interferometer implementing the cold ensemble exchange.} 
	(a) Front view: Two MOTs are loaded 36~mm apart. Cooling beams are shown in blue, probe beams in pink, and Raman beams in yellow. The trap is turned off, and the outer and inner cooling beams are blue and red detuned, respectively, which launches the ensembles towards each other. After the experiment, atoms are recaptured in the opposite trap to facilitate loading. (b) Side view: The design allows for four planes of optical access, enabling a compact apparatus. The vector  $\vec{g}$  shows the direction of gravity, while  $\vec{a}$  and  $\vec{\Omega}$  are the directions of acceleration and rotation measurement, respectively.
	(Figure and caption reproduced with permission from A. V. Rakholia \textit{et al}, Phys. Rev. Applied 2, 054012 (2014). Copyright 2014,  American Physical Society (Ref.~\cite{Rakholia2014})).}
	\label{fig:high_data_rate_rakholia}
\end{figure}

A first approach to increase the bandwidth of cold-atom sensor consists in reducing the interrogation time and operate with an efficient recapture of the atoms, as explored in several articles from the team at Sandia National Laboratories , following Ref.~\cite{McGuinness2012}. In particular, a dual-axis high-data-rate atom interferometer via cold ensemble exchange was reported in \cite{Rakholia2014}  (dual-axis accelerometer and gyroscope). By recapturing the atoms after the interferometer sequence, the authors maintained a large atom number at high data rates of 50 to 100 measurements per second. Two cold ensembles were formed in trap zones located a few centimeters apart and were launched toward one another (see see Fig.~\ref{fig:high_data_rate_rakholia}). During their ballistic trajectory, they were interrogated with a stimulated Raman sequence, detected, and recaptured in the opposing trap zone. The achieved  sensitivities were at 10$^{-5}$~m.s$^{-2}.\tau^{-1/2}$ and $\mu$\radps.$\tau^{-1/2}$ levels. 
Keeping the interest of accuracy, this approach combines both increase of bandwidth and reduction of size of the senors at the cost of sensitivity, leading to the possibility of a compromise depending of the application.

A more drastic approach is to avoid the production of the cold-atom ensembles by realizing atom interferometry in a  vapor cell as demonstrated in \cite{Biedermann2017}. This experiment, realized in Sandia National Laboratories, showed that interference signals may be obtained without laser cooling, by benefiting from  the Doppler selectivity of the atom interferometer resonance. With a data rate of 10~kHz and an interrogation time of $15 \ \mu$s, an inertial sensitivity of $10^{-1}$m.s$^{-2}.\tau^{-1/2}$ is demonstrated, with the prospect to improve the sensitivity by 2 orders of magnitude in the future. Although the proposed scheme is much simpler than a cold-atom sensor, the sensitivity  is still far from being competitive with that of best MEMS accelerometers than can reach sensitivities in the $10^{-6}$~m.s$^{-2}.\tau^{-1/2}$ range  (see e.g. Ref.~\cite{middlemiss_measurement_2016} and references therein).

\subsubsection{Hybridization with a classical accelerometer}

The heading of this subsection refers to the general motivation for a cold-atom based sensor: its inherent long-term stability. Nevertheless, the dead times in cold-atom sensors (associated with cold-atom preparation and detection) leads to a loss of inertial information and strongly mitigates the possibility to realize inertial measurement units (IMUs) based on atom interferometry, as pointed out  in Ref.~\cite{Jekeli2005}. Moreover, when keeping long interrogation times for high sensitivity, the vibrations lead to shot to shot interferometer phase fluctuations much higher than $2\pi$. 
The correlation with classical sensors during the interferometer measurement allows to lift the $\pi$ ambiguity in the phase determination (the atomic sensor being the fine scale of the vernier) and to improve the signal to noise ratio. This idea was first  demonstrated for a gravimeter \cite{Merlet2009} and later for an accelerometer in a plane \cite{geiger_detecting_2011}.
The hybridization technique develops the idea to measure the acceleration during the dead times of the atomic sensor. In a similar way as an atomic clock can steer a local oscillator (e.g. a quartz or a laser) to constrain its instability on several days, a cold-atom accelerometer can be used to servo the bias of a relative accelerometer featuring a larger bandwidth but a poorer bias stability \cite{lautier_hybridizing_2014,Cheiney2018}.

Ref.~\cite{lautier_hybridizing_2014} demonstrated at SYRTE  the concept of a hybrid accelerometer that benefits from the advantages of both conventional and atomic sensors in terms of bandwidth (DC to 430~Hz) and long term stability. The use of a real time correction of the atom interferometer phase by the signal from the classical accelerometer enabled to run it at best performance without any isolation platform, while a servo-lock of the DC component of the conventional sensor output signal by the atomic one realized the hybrid sensor. 
Following this work, a team at LP2N realized a navigation-compatible hybrid accelerometer using a Kalman filter where an algorithm was hybridizing the stable cold-atom interferometer with a classical accelerometer~\cite{Cheiney2018}. In particular, the bias of the classical accelerometer was tracked by the cold-atom sensor in an experimentally simulated harsh environment representative of that encountered in mobile sensing applications.  The resulting sensor exhibited a 400~Hz bandwidth and reached a stability of $10^{-7}$m.s$^{-2}$  after 11~h of integration.

\subsubsection{Multi-signal atom interferometers}

In the context of onboard applications with high dynamic range and high sensitivity, and in order to overcome the limit from the ambiguity in phase determination and the limit to the sensitivity when the atomic phase shift is closed to a multiple of $\pi$ rad, different propositions of multi-signal atom interferometer have been demonstrated.

%Overcoming dead times and increasing the robustness of cold-atom accelerometers can also be achieved using a multi-species sensor, as argued in Ref.~\cite{Bonnin2018}.
A first work analyzes configurations for improving the measurement range and sensitivity by relying on multi-species atom interferometry at ONERA, involving the simultaneous manipulation of different atomic species in a unique instrument to deduce inertial measurements~\cite{Bonnin2018}. Using a dual-species atom accelerometer manipulating simultaneously both isotopes of rubidium, the authors report a preliminary experimental realization of original concepts involving the implementation of two atom interferometers, first, with different interrogation times and, second, in phase quadrature. 

Two other experiments at the Weizmann Institute of Science have been achieved in the same context of operating cold-atom interferometers in mobile environments. First, a technique producing multiple phase measurements per experimental cycle was presented in Ref.~\cite{Yankelev2019}, allowing to realize quadrature phase detection in the presence of large phase uncertainties, and real-time systematic phase cancellation. Second, Ref.~\cite{Avinadav2019} introduces a scheme that improves on the trade-off between high sensitivity and limited dynamic range by a factor of 50 using composite fringes, obtained from sets of measurements with slightly varying interrogation times. The authors analyze  the performance gain in this approach and the trade-offs it entails between sensitivity, dynamic range, and temporal bandwidth.
%By combining composite-fringe measurements with a particle-filter estimation protocol, we demonstrate continuous tracking of a rapidly varying signal over a span two orders of magnitude larger than the dynamic range of a traditional atom interferometer.

 \subsubsection{Multi-axis measurements}
 
In order to realize an IMU, a 3-axis accelerometer and a 3-axis gyroscope is required. Even if multi-axis measurements have already been demonstrated, the possibility of measuring all components of inertia or even the 3 components of acceleration relies on successive measurements over the three directions. A theoretical proposal to extract information along several axis in a single shot was put forward at LP2N, where the authors  propose new multidimensional atom optics that can create coherent superpositions of atomic wave packets along three spatial directions~\cite{Barrett2019}. They argue how these tools can be used to generate light-pulse atom interferometers that are simultaneously sensitive to the three components of acceleration and rotation, and how to isolate these inertial components in a single experimental shot.

%\textcolor{red}{A Novel Monitoring Navigation Method for Cold Atom Interference Gyroscope \cite{Zhang2019}. Give performances of other systems, eg MEMS. Explain the integration efforts.} 

\subsubsection{Ways forward}
A large part of the works in developing atom accelerometers for mobile applications aimed at mitigating the problem of phase ambiguity and loss of information associated with dead times.
The high data rate interferometers using recapture methods partially reduce the problem of dead times but at the cost of strong reduction of sensitivity. On the other hand, the inertial noise aliasing associated with dead times can be alleviated by the hybridization technique, but this method is limited by non linearity in the correlation between both sensors. To solve this problem, combining different approaches will probably be needed. As an example, combining hybridization with continuous operation (i.e. without dead times) and eventually interlevead measurements, already demonstrated for rotation measurements\cite{Dutta2016,Savoie2018}, should enable to achieve the full potential (i.e. quantum limited sensitivity) of cold-atom accelerometers.

%\subsubsection{Alternative approaches: trapped atoms or atomic vapor cell}
%We briefly mention here two alternatives to the strategy presented in the previous paragraph (interrogation of free falling cold-atom clouds to realize a sensor with a good long-term stability). These developments are triggered by the wish to realize more integrated sensors or simplified schemes without laser cooling, at the cost of a degradation of the stability.

%Ref.~\cite{Akatsuka2017} demonstrates an atom interferometer operating on the $^1S_0–^3P_0$ clock transition of $^{87}$Sr atoms in a magic optical guide, where the light shift perturbations of the guiding potential are canceled. As a proof-of-principle demonstration, a Mach–Zehnder interferometer is set horizontally to map the acceleration introduced by the focused optical guide.  Possible applications of such a magic guide interferometer, including a hollow-core fiber interferometer and gradiometer, are discussed.
%Subsequently, an atom interferometer inside a hollow-core photonic crystal fiber was reported in \cite{Xin2018}, where the use of the optical fields in a hollow-core photonic crystal fiber enabled to spatially split, reflect, and recombine a coherent superposition state of free-falling $^{85}$Rb atoms to realize an accelerometer. However, the  interrogation times explored were limited to $100 \ \mu$s, yielding a poor inertial sensitivity for the moment, compared to other highly integrated technologies (e.g. MEMS accelerometers).

\section{\label{sec:other_measurements} Other measurements of inertial effects}
%\textcolor{red}{\textit{Shall we list here briefly the other experiments ? Recoil measurements, trapped atom interferometers, tests of equivalence principle, gravitational wave detection.}}

The sensitivity of atom interferometers to inertial forces can also be exploited for precise measurements of fundamental constants and fundamental tests, for the search for new exotic forces, for dark matter detection, for gravitational wave detection. For completeness of this review, we will briefly mention in this section some of these applications, and refer to the review articles cited in the introduction for further details.

Atom interferometry is for instance key to precisely measure the change of velocity undergone by an atom after the transfer of momenta of a large number of photons \cite{Cadoret2008}. This is accomplished by using a Ramsey Bord\'e interferometer\cite{Borde1984}, which acts as a sensor for the velocity change between the first and second part of the interferometer, both constituted of a pair of $\pi/2$ pulses. This change of velocity is realized by placing the atoms, in between the two pairs of pulses, in an accelerated lattice. There the atoms undergo a large number of Bloch oscillations (up to a thousand), which results in a momentum transfer of $N\hbar k$. This allows for a precise determination of the recoil velocity. Remarkably, and this is the main motivation for such an experiment, this allows for the determination of fine structure constant $\alpha$, the dimensionless constant that characterizes the strength of electromagnetic interactions, with relative uncertainties below the $10^{-9}$ level \cite{Bouchendira2011, Parker2018}. This experimental determination can finally be compared with the more indirect determination of $\alpha$ derived from the anomalous magnetic moment of the electron, which can be precisely calculated using quantum electrodynamics (QED) theory and out of which a value the fine structure constant can be determined. The comparison between the two determinations is one the most stringent tests of QED today.

Another prospected application of cold-atom inertial sensors is the detection of gravitational waves \cite{dimopoulos_atomic_2008,Geiger2016}. In the currently mostly considered schemes, a set of atom accelerometers  placed along a very long baseline (of hundreds of meters, if not kilometers) are simultaneously interrogated with common laser beamsplitters, in a gradiometer-like measurement configuration \cite{Canuel2018,Zhan2019}. This will allow for the detection of gravitational waves, whose signature is actually identical to gravity gradients, in a so far unexplored frequency band for ground-based detectors (0.1-1~Hz). Several studies have also proposed space-based detectors to address the mHz frequency band (e.g. \cite{Graham2016a,AEDGEarxiv}).

Finally, other tests of gravitational physics can be performed, such as tests of the Weak Equivalence Principle, by comparing the acceleration felt by two different atomic species \cite{ Fray2004,Bonnin2013,Schlippert2014,Tarallo2014, Zhou2015,Barrett2016,Duan2016,Rosi2017}. To push the relative accuracy of these atomic physics based tests  below the current limits of experiments involving classical test masses (at the $10^{-14}$ level\cite{Touboul2017}), long interrogation times are required. On ground, dedicated facilities are being operated or currently  built, where atoms will be launched or dropped over a few seconds. This can be realized in very tall vacuum chambers of typically 10 meters \cite{Sugarbaker2013}, or even longer, as well as in zero-g simulators, such as drop towers \cite{VanZoest2010} or in zero-g planes \cite{Barrett2016}. This prepares the ground for future space missions, such as the STE-QUEST mission \cite{Altschul2015}, where interferometers would last tens of seconds, thus boosting the sensitivity  by 3 to 4 orders of magnitude.

\section{New techniques for cold-atom interferometry }
\label{sec:new_techniques}

Though the domain of cold-atom interferometry has gained a considerable maturity, leading, for example, to the industrial transfer of the technology, the  performances of these instruments can still be improved significantly for various applications. In the recent years, new methods have been introduced and are still presently being actively investigated.

\subsection{Large Momentum Transfer (LMT) atom optics}

A variety of advanced beamsplitting methods have been demontrated such as double Raman transitions \cite{Leveque2009, Malossi2010} and double Bragg diffraction \cite{ahlers_double_2016}, high order Bragg diffraction \cite{muller_atom_2008}, sequences of Bragg pulses \cite{chiow_102_2011}, Bloch oscillations in accelerated optical lattices \cite{Clade2009}, combination of Bloch oscillation and high order Bragg pulses \cite{muller_atom_2009, mcdonald2013, mcdonald_faster_2014}, frequency-swept rapid adiabatic passage \cite{kovachy_2012,kotru_large-area_2015}. All these methods allow for improving the scale factor of the sensors, through a drastic increase of the separation of the atomic wavepackets, which can reach about a hundred photon momentum \cite{chiow_102_2011, kovachy_quantum_2015, Plotkin-Swing2018}, or even more. Record separation and subsequent recombination with up to 408 photon momentum separation has recently been demonstrated \cite{gebbe2019twinlattice}.

A drawback of these methods lie in the existence of diffraction phases, which are parasitic phase shifts due to the beamsplitting process that depend on the depth of the lattice potential \cite{Buechner2003, kovachy_optical_2010}, and to the presence of parasitic interferometers due to the multiport nature of Bragg diffraction \cite{parker_controlling_2016}. This leads to systematics in the interferometer phase, but also to phase noise due to intensity temporal fluctuations, and loss of contrast, induced by intensity inhomogeneities. Strategies are being developed to mitigate these side effects \cite{Estey2015, Gochnauer2019}. This explains why, although a number of proofs of principle have been made, only few instruments have demonstrated a clear gain on the measurement of an actual inertial quantity. In particular, the  team at Stanford University implemented these techniques in an atomic gradiometer \cite{asenbaum_phase_2017}. With transitions of 20 $\hbar$k, they obtained a gradient sensitivity of $3\times10^{-9}\text{s}^{-2}$ per shot (but with a very long cycle time of 15~s). The differential phase noise was  relatively large though, of about 130~mrad per shot for  transitions of 30 $\hbar$k, which is well above the detection noise limit and leaves margin for significant improvement in the future.

 In the context of large-momentum transfer beam splitters which require larger laser intensities, schemes to interface an atom interferometer with an optical cavity tuned at resonance have been explored \cite{Hamilton2015,RiouMielec2017,Dovale-Alvarez2017,Sapam2020inprep}. Such schemes are appealing since an optical resonator can provide an interferometric control of the mode and  important laser power enhancement ($\gg 100$). Nevertheless, the beam size that must be reached in the resonator (several mm of waist) to keep an homogeneous intensity profile over the freely expanding atom cloud tends to push towards long (several meters) cavities \cite{Dovale-Alvarez2017} or towards the degenerate regime \cite{RiouMielec2017,Sapam2020inprep} where the control of the impact of optical aberrations becomes challenging. 
Reducing the impact of intensity inhomogeneities over the atomic cloud can also be achieved by increasing the size of gaussian beams (at the cost of optical power loss) \cite{asenbaum_phase_2017} or by using such as so-called top-hat beams with a flat intensity profile \cite{Mielec2018}.

\subsection{Ultracold atom sources}
Another important axis of investigation is the improved control over the atomic source, offered by ultracold atoms. Their reduced ballistic expansion allowed for increasing the interrogation time \cite{kovachy_quantum_2015}, and their narrow momentum distribution for improving the efficiency of LMT methods \cite{Szigeti2012}. More, lower temperatures also reduce systematic effects, such as related to the exploration by the atoms of intensity inhomogeneities in the spatial profile of the beamsplitters, or to wavefront distortions and Coriolis accelerations. Lensing methods, such as Delta Kick collimation in atom chips \cite{ muntinga_interferometry_2013} and optical traps \cite{ kovachy_matter_2015}, allowed for the production of well collimated source, with temperatures lower than 100 pK. New detection methods have been demonstrated which allow for spatially resolving the variations of the interferometer phase across the atomic source, at the output of the interferometer \cite{Dickerson2013,Sugarbaker2013}, increasing the fringe visibility and the dynamical range of the sensors.

\subsection{Alkaline-Earth atoms}

Finally, other atom sources are also being used, such as other alkali species (eg. Potassium for tests of the Weak Equivalence Principle), or alkaline-earth atoms, such as Yb \cite{hartwig_testing_2015} or Sr \cite{ del_aguila_bragg_2018}. The latter offer, for their bosonic isotopes having zero spin in the ground state, reduced sensitivity to stray magnetic fields, and a richer electronic structure, with narrow lines that can be used to implement single photon beamsplitters \cite{hu_atom_2017,Rudolph_large_2020} or to implement quantum metrology measurement protocols, such as based on spin-squeezing \cite{ hosten_measurement_2016, salvi_squeezing_2018}.

\section{Conclusion}
Cold-atom inertial sensors have been developed for nearly 30 years in several groups worldwide, with an acceleration of the research effort and of the industrial transfer  in the last 10 years. These sensors are especially well suited for applications requiring high performance in terms of stability and accuracy, while they comparatively currently feature a weak robustness, dynamic range  and level of miniaturization. Therefore, they have found up to now natural applications in testing fundamental physics, in metrology and in geoscience. Nevertheless, there is a growing research effort on taking the technology out of the laboratory, in particular on  realizing high accuracy inertial measurement units operating in mobile platforms. 
On the fundamental side, several experiments are being set up to look for new physics, and large-scale instruments for gravitational wave detection are  under design and realization in 3 continents.

Several atom interferometer architectures have been considered with different atom optics techniques and different atomic species. So far, the best performances have been achieved by cold-atom sensors using alkali atom sources and two-photon stimulated Raman transitions. Several teams are working on  techniques to enhance the sensitivity and accuracy, in particular on ways to increase the separation between the two arms of the interferometer, using colder atom samples or atomic species with a richer level structure. Therefore, much improvement in performance can still be expected from this technology.

Many groups are also working on simplifying the sensor architecture and subsystems to broaden the scope of applications and making cold-atom sensors compatible with operation in the field. On the industrial side, several companies have started to tackle the challenge of integrating the subsystems and currently concentrate on developing gravimeters or accelerometers with cold Rb atom, for which efficient, robust, and qualified laser sources have been realized. These efforts will ease the deployment of atom interferometer sensors and allow them to address a wide range of new applications, from ground to space, beyond the reach of classical sensors.

\section*{\label{sec:summary_points} Summary points} 
\begin{enumerate}
\item Atom interferometers use quantum superpositions  of different momentum states in atoms.
\item Realizing these superposition states is efficiently achieved by using the interaction of an atom with two counter-propagating laser beams, which realize the role of beam splitter and mirror for the atomic wave.
\item The scale factor of atom interferometers (link between the inertial effect to be measured and the interferometer output phase) is given by the space-time area of the interferometer, which is proportional to the square of the time spent by the atoms in the interferometer and to the wave-vector of the interrogation lasers.
\item Using cold-atom sources for atom interferometry allows to reach larger scale factors and high levels of stabilitiy and accuracy. Such performance originate  from the good control of atomic trajectories associated with low temperatures.
\item The fundamental limit to the sensitivity in these sensors is the quantum projection noise associated with the projective measurement of the atomic state performed at the output of the interferometer for ensembles of typically 1 million of atoms. Nevertheless, this quantum noise limit  is often not reached since  the effect of vibration noise  dominates. 
\item The mostly employed atoms are Rubidium and Cesium (Potassium in few cases), cooled by lasers to a temperature of few micro-Kelvins. Interferometers with alkaline-Earth atoms (Strontium and Ytterbium) attract more and more interest. 
\item Cold-atom inertial sensors mainly consist of a vacuum chamber hosting the atom source, optical systems to realize the momentum state superpositions, light detectors, real-time control electronics   and external instruments such as mechanical accelerometers. Typical dimensions of instruments range from 10~cm (integrated versions for field applications) to 10~m (large-scale experiments for fundamental studies). 
\item Gravimeters are the most studied atom interferometers. Best instruments have inaccuracy better than 2$\times10^{-8}~$ m.s$^{-2}$ and reach stability of 5$\times10^{-10}~$m.s$^{-2}$ in $10^5~$s of measurement. These sensors were the first  instruments to be commercially developed.
\item Other largely studied sensors are accelerometers, rate gyroscopes and gravity gradiometers. Few large scale infrastructures (over 100~m baselines) are developed as prototypes of gravitational-wave detectors in the deci-Hertz frequency band.
\item Owing to their stability and accuracy, cold-atom intertial sensors have natural applications in geosciences and tests of fundamental physics. Integration and engineering efforts let anticipate important applications in strategic inertial navigation in a near future. 
\item Several research group work on new techniques to improve the sensitivity, stability, accuracy or compactness of cold-atom inertial sensors. Particularly followed routes are implementations of large momentum transfer beam splitters, production of ultracold-atom sources within a short time or new optical systems for improved atom optics efficiency.
\item About 50 research groups (including around 7 private companies) are active in the field.
\end{enumerate}

\begin{acknowledgments}
We acknowledge Quentin Beaufils for careful reading of the manuscript.
\end{acknowledgments}

\section*{Data Availability Statement} 
The data that support the findings of this study are available from the corresponding author upon reasonable request.

\section{References}
%\nocite{*}
%\bibliography{biblio_review_AVS}% Produces the bibliography via BibTeX.

%

\appendix
%\section{Appendixes}

\section{\label{sec:} List of groups working on cold-atom interferometers}
\label{sec:list_groups}

\begin{figure*}[h]
	\centering
	\includegraphics[width=\linewidth]{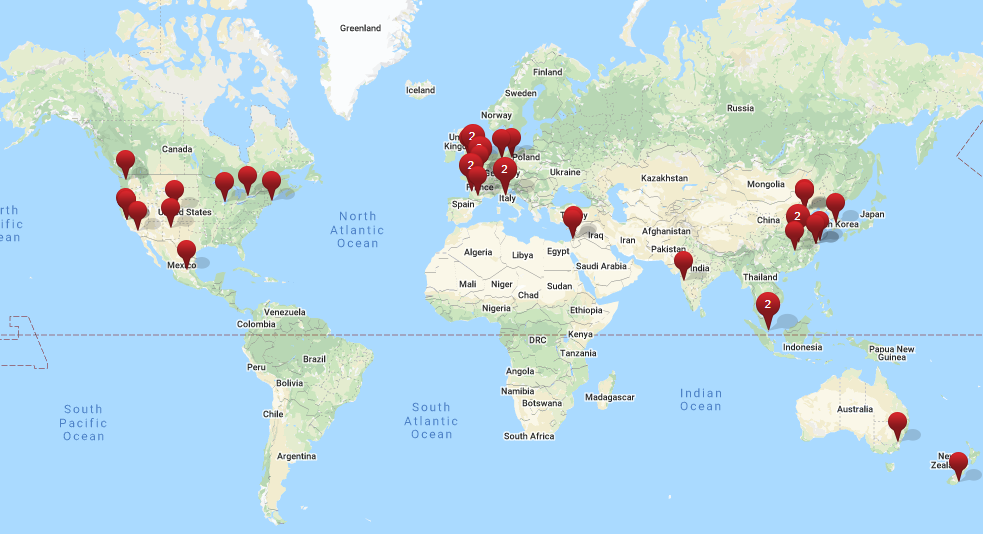}
	\caption{\textbf{World-map with the different research groups and companies active in the field of cold-atom intertial sensors}. List of groups  as described in Tables \ref{tab:context1},\ref{tab:context2},\ref{tab:context3}. Map realized with \url{https://fortress.maptive.com/}.}
	\label{fig:world_map_groups}
\end{figure*}

The tables below (\ref{tab:context1},\ref{tab:context2},\ref{tab:context3}) summarize the research effort of the main actors in the field and Fig.~\ref{fig:world_map_groups} shows a map with the different groups.
%\textcolor{red}{Check that we do not miss a group. People will have some time (e.g. 1 week to give feedback after our arxiv submission.}

\begin{table*}[!t]
\centering
\begin{tabular}{|l|p{3cm}|p{7cm}|p{3cm}|l|}
\hline
\textbf{country} & \textbf{institution} &  \textbf{expertise (atomic species)}  & \textbf{link} \\
\hline
\hline
France & SYRTE, Paris &  gravimeter, gyroscope, gradiometer, trapped AI, GW detection, atom chips  (Rb, Cs) & \web{https://syrte.obspm.fr/spip/science/iaci/?lang=en} \\
\hline
France & LKB, Paris & $h/m$, LMT, Bloch (Rb) & \web{http://www.lkb.upmc.fr/metrologysimplesystems/atom_interf/} \\
\hline
France & ONERA, Palaiseau &  gravimeter, gradiometer, WEP test, field applications (Rb) & \web{https://www.onera.fr/en/news/shom-onera-cold-atoms-gravimetry} \\
\hline
France & LP2N, Bordeaux &  WEP test, GW  (Rb, K, Sr) & \web{https://www.coldatomsbordeaux.org/}\\
\hline
France & LCAR, Toulouse Univ. &  test of atom neutrality, topological phases, atom chips (Rb)  & \web{https://www.quantumengineering-tlse.org/research/atom-interferometry/} \\
\hline
France & \textit{Muquans}, Bordeaux &  gravimeter, gradiometer (Rb) & \web{https://www.muquans.com/} \\
%\hline
%France & \textit{Thales}, Palaiseau &  accelerometer, atom chip (Rb) & \web{https://www.thalesgroup.com/en/global/innovation/research-and-technology} and Ref.~\cite{Dupont-Nivet2018}\\
\hline
France & \textit{iXBlue}, Bordeaux &  inertial sensors  (Rb) & \web{https://www.coldatomsbordeaux.org/ixatom} \\
\hline
\hline
Germany & Hannover Univ. & gravimeter, gradiometer, WEP test, GW detection, LMT,  atom chips (Rb, K, Yb) & \href{https://www.iqo.uni-hannover.de/de/arbeitsgruppen/quantum-sensing/}{website}\\
\hline
Germany & Humboldt Univ. &  gravimeter, EP  (Rb) & \href{https://www.physics.hu-berlin.de/en/qom/research}{website} \\
%\hline
%\hline
%Greece & IESL-FORTH, Crete &		guided interferometry, BEC 	(Rb) & \web{https://www.iesl.forth.gr/en/research/bec} \\
\hline
\hline
Italy & LENS, Florence &  gradiometer, trapped AI, WEP test  (Rb, Sr, Cd) & \web{http://coldatoms.lens.unifi.it/} \\
\hline
Italy & \textit{AtomSensors} & gravimeter, gradiometer  (Rb) & \web{http://www.atomsensors.com/index.php/en/}\\
\hline
\hline
Israel & Weizman Institute of Science, Rehovot   &		gravimeter, inertial navigation  (Rb) & \web{http://www.weizmann.ac.il/complex/Firstenberg/research-activities/atom-interferometry}\\
\hline
Israel & \textit{Rafael Ltd} &  inertial navigation unit (Rb) & \web{https://www.rafael.co.il/} and Ref.~\cite{Yankelev2019} \\
\hline
\hline
UK & Univ. Birmingham &  gradiometer, field applications  (Rb) & \web{https://www.birmingham.ac.uk/research/activity/gravity/index.aspx}\\
\hline
UK & Imperial College, London &		accelerometer, search for dark energy 	 (Rb) & \web{http://www.imperial.ac.uk/centre-for-cold-matter/research/cold-atoms/}\\
%\hline
%UK & Univ. Nottingham &		guided AI  (Rb) & \web{https://coldatomsgroupnottingham.com/}\\
\hline
UK & \textit{Teledyne e2v} &  gravimeter  (Rb) & \web{https://www.teledyne-e2v.com/products/quantum-technology/our-key-projects/}\\
\hline
UK & \textit{M2 lasers} &  accelerometer  (Rb) & \web{http://www.m2lasers.com/quantum.html}\\
\hline
\hline

\end{tabular}
\caption{Table of the main actors in the field of  inertial sensors based on free-falling cold atoms (\textbf{Europe and EU-affiliated area}). AI: Atom interferometry; WEP: Weak Equivalence Principle; GW: Gravitational Wave; LMT: Large Momentum Transfer techniques; BEC: interferometry with Bose Einstein Condensates; $h/m$: measurement of the recoil velocity. \textit{Companies are shown in italics.}}
\label{tab:context1}
\end{table*}

\begin{table*}[!h]
\centering
\begin{tabular}{|l|p{3cm}|p{7cm}|p{3cm}|l|}
\hline
\textbf{country} & \textbf{institution} &  \textbf{expertise (atomic species)}  & \textbf{link} \\
\hline
\hline
Canada & York University & echo interferometry  (Rb) & \web{https://www.physics.yorku.ca/faculty-profiles/kumarakrishnan-anantharaman/}\\
\hline
\hline
Mexico & Univ. San Lui Potosi & gravimeter  (Rb) & \web{https://www.ifisica.uaslp.mx/~egomez/projects/research.html}\\
\hline
\hline
USA & Stanford Univ. &  WEP test, GW detection, LMT, BEC, 10-meter fountain (Rb, Sr) & \href{https://web.stanford.edu/group/kasevich/cgi-bin/wordpress/}{website} \\
\hline
USA & UC Berkeley &  tests of fundamental physics, $h/m$, LMT, gravimeter  (Cs, Li) & \href{http://matterwave.physics.berkeley.edu/}{website} \\
\hline
USA & JPL, Pasadena & applications in geodesy,  gradiometry (Rb) & \href{https://scienceandtechnology.jpl.nasa.gov/people/n_yu}{website} \\
\hline
USA & Sandia National Lab., Albuquerque &  high sampling rates, multi-axis, vapor cell (Cs, Rb) & \href{https://www.sandia.gov/mstc/quantum/index.html}{website} \\
\hline
USA	& Draper Lab, Cambridge &  LMT (Cs) & Ref.~\cite{kotru_large-area_2015} \\
\hline
USA & Univ. Washington &  LMT, BEC  (Yb) & \href{http://faculty.washington.edu/deepg/}{website}\\
\hline
USA & \textit{AO Sense Inc.} &  gravimeter, inertial sensors  (Rb) & \href{https://aosense.com/}{website} \\
\hline
USA & NIST, Boulder &  miniature AI for inertial sensing  (Rb) & \href{https://www.nist.gov/people/john-kitching}{website} \\
\hline
USA & Northwestern Univ. & GW detection, LMT   & \href{http://faculty.wcas.northwestern.edu/tim-kovachy/}{website} \\
\hline
USA & Goddard (NASA) & gradiometer  & \href{http://absimage.aps.org/image/DAMOP17/MWS_DAMOP17-2017-000729.pdf}{link} \\
\hline
\hline

\end{tabular}
\caption{Table of the main actors in the field of  inertial sensors based on free-falling cold atoms (\textbf{North America}). \textit{Companies are shown in italics.}}
\label{tab:context2}
\end{table*}

\begin{table*}[!t]
\centering
\begin{tabular}{|l|p{3cm}|p{7cm}|p{3cm}|l|}
\hline
\textbf{country} & \textbf{institution} &  \textbf{expertise (atomic species)}  & \textbf{link} \\
\hline
\hline
Australia &	ANU, Canberra	&  gravimeter, gradiometer  (Rb) & \web{http://atomlaser.anu.edu.au/AQSG_research.html}\\
\hline
\hline
China & WIPM, Wuhan 	& gravimeter, gyroscope, EP, GW  (Rb, Sr, Cs) 10-meter fountain & \web{http://english.wipm.cas.cn/rh/rd/yzfzsys/yzfz1/yzfz1_res/}\\
\hline
China & Zhejiang Univ. and Zhejiang Univ. of Technology, Hangzhou & gravimeter, gradiometry  (Rb) & \cite{Fu2019} \\
\hline
China & Zhejiang Univ., Hangzhou & gravimeter  (Rb) & \web{https://doi.org/10.1088/1555-6611/aafd26} \\
\hline
China & HUST, Wuhan &  gravimeter, gyroscope, EP  (Rb) & \web{https://www.researchgate.net/scientific-contributions/33262158_Zhong-Kun_Hu}\\
\hline
China & NIM, Beijing &  gravimeter  (Rb) & \cite{Wang2018}\\
\hline
China & CIMM, Beijing & gravimeter  (Rb) & \web{http://www.cimm.com.cn}\\
\hline
China & USTC, Shanghai & gravimeter  (Rb) & \web{http://quantum.ustc.edu.cn/web/index.php/en/node/258} \\
\hline
China & NUDT, Changsha & gravimeter  (Rb) &  \\
%\hline
%China & Sun Yat-Sen University, Zhuhai & Trapped gravimeter (Rb) & \cite{YongguanKe_compact_2018} \\
\hline
China & Tsinghua University, Beijing & cold atom beam gyroscope  (Rb) &  \\
\hline
\hline
Korea & KRISS, Daejon &  gravimeter  (Rb) & \href{http://aappsbulletin.org/myboard/down.php?Board=featurearticles&filename=Vol27_No1_Feature%20Articles-2.pdf&id=185&fidx=1}{article}\\
\hline
\hline
India & IISER Pune &  AI with BEC  (Rb) & \web{http://www.iiserpune.ac.in/~umakant.rapol/research/}\\
%\hline
%\hline
%Japan & Univ. Tokyo	&  optically guided interferometry  (Sr) & \cite{Akatsuka2017} \\
\hline
\hline
New Zealand & Univ. Otago &  gravimeter  (Rb) &\web{http://www.physics.otago.ac.nz/nx/mikkel/atom-interferometer.html} \\
\hline
\hline
Singapore	& CQT &	 gravimeter  (Rb) & \web{http://www1.spms.ntu.edu.sg/~rdumke/expc1.html}\\
%\hline
%Singapore & NTU &   hollow core fiber	AI  (Rb) & \web{http://www1.spms.ntu.edu.sg/~langroup/index.html}\\
\hline
Singapore & \textit{Atomionics} &  inertial sensors  & \web{http://www.atomionics.com/}\\

\hline
\hline

\end{tabular}
\caption{Table of the main actors in the field of  inertial sensors based on free-falling cold atoms (\textbf{Asia and Oceania}). \textit{Companies are shown in italics.} }
\label{tab:context3}
\end{table*}

\end{document}